\documentclass[]{aa}
\usepackage{graphicx}
\newcommand{\kmprs}  {\mbox{\rm km\,s$^{-1}$}}

\newcommand{\feh} {\mbox{\rm [Fe/H]}}
\newcommand{\znh} {\mbox{\rm [Zn/H]}}
\newcommand{\sh} {\mbox{\rm [S/H]}}
\newcommand{\sfe} {\mbox{\rm [S/Fe]}}
\newcommand{\szn} {\mbox{\rm [S/Zn]}}

\newcommand{\znfe} {\mbox{\rm [Zn/Fe]}}
\newcommand{\mgfe} {\mbox{\rm [Mg/Fe]}}
\newcommand{\alphafe} {\mbox{\rm [$\alpha$/Fe]}}
\newcommand{\teff}  {\mbox{$T_{\rm eff}$}}
\newcommand{\logteff} {\mbox{${\rm log}\,T_{\rm eff}$}}
\newcommand{\logg}  {\mbox{{\rm log}\,$g$}}
\newcommand{\HI} {\ion{H}{i}} 

\newcommand{\OI} {\ion{O}{i}} 
\newcommand{\SI} {\ion{S}{i}} 
\newcommand{\SII} {\ion{S}{ii}} 
\newcommand{\MgI} {\ion{Mg}{i}} 
\newcommand{\MgII} {\ion{Mg}{ii}} 
\newcommand{\CaI} {\ion{Ca}{i}} 

\newcommand{\FeI} {\ion{Fe}{i}}
\newcommand{\FeII} {\ion{Fe}{ii}}
\newcommand{\CrII} {\ion{Cr}{ii}}
\newcommand{\ZnI} {\ion{Zn}{i}}
\newcommand{\ZnII} {\ion{Zn}{ii}}

\newcommand{\Mv} {\mbox{$M_V$}}

\newcommand{\mvphot} {\mbox{$M_{V,{\rm phot}}$}}
\newcommand{\mvpar} {\mbox{$M_{V,{\rm par}}$}}

\newcommand{\water}{\mbox{\rm H$_2$O}}
\newcommand{\ratfeh} {\mbox{${\rm [\frac{Fe}{H}]}$}}
\newcommand{\ratsh} {\mbox{${\rm [\frac{S}{H}]}$}}
\newcommand{\ratznh} {\mbox{${\rm [\frac{Zn}{H}]}$}}
\newcommand{\ratsfe} {\mbox{${\rm [\frac{S}{Fe}]}$}}
\newcommand{\ratznfe} {\mbox{${\rm [\frac{Zn}{Fe}]}$}}
\newcommand{\ratszn} {\mbox{${\rm [\frac{S}{Zn}]}$}}
\def\ltsima{$\; \buildrel < \over \sim \;$}
\def\simlt{\lower.5ex\hbox{\ltsima}}
\def\gtsima{$\; \buildrel > \over \sim \;$}
\def\simgt{\lower.5ex\hbox{\gtsima}}
\begin{document}

\title{Sulphur and zinc abundances in Galactic stars and damped Ly$\alpha$ systems
\thanks{Based on observations collected at the European
Southern Observatory, Chile (ESO No. 67.D-0106)}}


\author{P.E.~Nissen \inst{1} \and Y.Q.~Chen \inst{2}
\and M.~Asplund \inst{3} \and M. Pettini \inst{4}}

\offprints{P.E.~Nissen}

\institute{
Department of Physics and Astronomy, University of Aarhus, DK--8000
Aarhus C, Denmark. 
\email{pen@ifa.au.dk}
\and National Astronomical Observatories, Chinese Academy of Sciences,
   Beijing 100012, P.R. China.
\email{cyq@bao.ac.cn}
\and
Research School of Astronomy and Astrophysics,
Australian National University, Mount Stromlo Observatory,
Cotter Road, Weston, ACT 2611, Australia.
\email{martin@mso.anu.edu.au}
\and Institute of Astronomy, University of Cambridge, Madingley Road, Cambridge, CB3 0HA, UK.
\email{pettini@ast.cam.ac.uk}
}

\date{Received 9 July 2003 / Accepted November 19 2003}

\abstract{High resolution spectra of 34 halo population dwarf and subgiant stars
have been obtained with VLT/UVES and used to derive sulphur abundances from
the $\lambda \lambda 8694.0, 8694.6$ and $\lambda \lambda 9212.9, 9237.5$
\SI\ lines. In addition, iron abundances have been determined from 19 
\FeII\ lines and zinc abundances from the $\lambda \lambda 4722.2, 4810.5$
lines. The abundances are based on a classical 1D, LTE model atmosphere
analysis, but effects of 3D hydrodynamical modelling on the [S/Fe], [Zn/Fe]
and [S/Zn] ratios are shown to be small.  We find that most halo stars 
with metallicities in the range $-3.2 < \feh < -0.8$ have
a near-constant $\sfe \simeq +0.3$; a least square fit to \sfe\ vs. \feh\
shows a slope of only $-0.04 \pm 0.01$. Among halo stars with 
$-1.2 < \feh < -0.8$ the majority have $\sfe \simeq +0.3$, but two stars
(previously shown to have low $\alpha$/Fe ratios) have $\sfe \simeq 0.0$.
For disk stars with $\feh > -1$, \sfe\ decreases with
increasing \feh . Hence, sulphur behaves like other typical
$\alpha$-capture elements, Mg, Si and Ca. Zinc, on the other hand, traces 
iron over three orders of magnitude in \feh, although there is some 
evidence for a small systematic Zn overabundance ($\znfe \simeq +0.1$)
among metal-poor disk stars and for halo stars with $\feh < -2.0$. Recent 
measurements of S and Zn in ten damped Ly$\alpha$ systems (DLAs) with redshifts
between 1.9 and 3.4 and zinc abundances in the range $-2.1 < \znh < -0.15$
show an offset relative to the \szn\ -- \znh\ relation in Galactic
stars. Possible reasons for this offset are discussed, including
low and intermittent star formation rates in DLAs.
}

\maketitle

\section{Introduction}
\label{introduction}

Sulphur is generally regarded as an $\alpha$-capture element. The
work on Galactic stars by Fran\c{c}ois (\cite{francois87},
\cite{francois88}) supported this view by showing that [S/Fe]
increases from zero at solar metallicities to a plateau level
of about +0.5 dex in the metallicity range $-1.8 < \feh < -1.0$.
This is an analogous behaviour to those of other $\alpha$-elements,
Mg, Si, and Ca - see Norris et al. (\cite{norris01}) and Carretta et al.
(\cite{carretta02}), who both find \alphafe\ to be nearly constant at
a level of +0.4\,dex in the metallicity range $-4 < \feh < -1$
and then to decrease towards the solar \alphafe\ ratio for $\feh > -1$.
The standard interpretation is that this trend arises from the
time delay in the production of about two thirds of the iron by
supernovae (SNe) of Type Ia relative to the near-instantaneous release
of the $\alpha$-elements by Type II SNe. In this connection it should
be noted that the metallicity at which Type Ia SNe start to contribute
(the ``knee'' of the [$\alpha$/Fe] - [Fe/H] trend) seems to depend on
the orbital properties of the stars. Halo stars with $\feh \sim -1$
belonging to the outer halo tend to have lower \alphafe\ than stars
in the inner halo (Nissen \& Schuster \cite{nissen97};
Stephens \& Boesgaard \cite{stephens02}; Gratton et al. \cite{gratton03}).
Stephens \& Boesgaard find a slope of $-0.15$ for the mean \alphafe\
vs. \feh\ for 53 stars in the metallicity range $-3.8 < \feh < -0.6$,
but this is for a sample with special kinematics, i.e. stars with large
maximum distances from the Galactic center and/or the Galactic plane
or stars having extreme retrograde motion.

Recent observations of the $\lambda \lambda 8694.0, 8694.6$\,\AA\
\SI\ lines in spectra of  metal-poor stars by
Israelian \& Rebolo (\cite{israelian01}) have, however, challenged the view
that sulphur is an $\alpha$-element. Their data,
obtained with the 4-m William Herschel Telescope on La Palma,
suggest that \sfe\ increases linearly with decreasing \feh\ to a level
as high as $\sfe \sim +0.7$ at $\feh = -2.0$. The study of Takada-Hidai et al.
(\cite{takada02}), based on Keck High Resolution Echelle Spectrograph (HIRES)
observations, supports a quasi linear dependence of \sfe\ on \feh\
although in their case \sfe\ reaches only $+0.5$ dex at $\feh = -2.0$.

As a possible explanation of the high value of \sfe\ in metal-poor stars,
Israelian \& Rebolo (\cite{israelian01}) proposed that very massive 
supernovae with exploding He-cores and a high explosion energy
make a significant contribution to the early chemical evolution
of galaxies. According to Nakamura et al. (\cite{nakamura01}) these hypernovae
overproduce S with respect to O, Mg and Fe. With
this intriguing possibility in mind a more thorough investigation
of sulphur abundances in halo stars seems worthwhile.

A clarification of the trend of S abundances is also much needed
in deciphering the chemical enrichment of
damped Ly$\alpha$ systems (DLAs), widely regarded as the
progenitors of present-day galaxies at high redshift. The
importance of sulphur stems from the fact that, unlike most other heavy
elements, S is not depleted onto dust. Consequently, observations
of the relatively weak \SII\ triplet resonance lines at $\lambda\lambda
1250, 1253, 1259$\,\AA\ yield a direct measurement of the abundance of
S in DLAs. Another element for which this is the case is
Zn, and indeed most of our current knowledge of the chemical
evolution of the universe at high redshift is based on surveys of
[Zn/H] in DLAs (e.g. Pettini et al. \cite{pettini99};
Prochaska \& Wolfe \cite{prochaska02}). {\it If} S is an $\alpha$-capture
element, then its abundance relative to Zn (assumed to be an iron-peak element)
could be used as `a chemical clock' to date the star-formation
process at high $z$. Specifically, if a major star formation
episode in a DLA occurred within $\approx 0.5$\,Gyr 
prior to the time when we observe the galaxy, 
we would expect to measure an enhanced [S/Zn] ratio, and
{\it vice versa}.
Measurements of [S/Zn] in DLAs have been relatively scarce
until recently, but are now becoming available in increasing numbers
thanks the Ultraviolet and Visual Echelle Spectrograph
(UVES) on the ESO Very Large Telescope (VLT).
The VLT/UVES combination affords high spectral resolution and
high efficiency over most of the optical spectrum,
from blue and near-UV wavelengths to the far red.
However, without a secure
knowledge of the behaviour of S in metal-poor Galactic stars 
we clearly stand little chance of interpreting the situation at high $z$.

In the present paper we report on a survey of sulphur abundances in 34
metal-poor dwarf stars based on high resolution observations of
the $\lambda \lambda 8694.0, 8694.6$ and
$\lambda \lambda 9212.9, 9237.5$\,\AA\ \SI\ lines 
obtained with UVES.
The $\lambda \lambda 4722.2, 4810.5$\,\AA\
\ZnI\ lines are also included in our spectra, allowing us to study the Galactic
evolution of both sulphur and zinc.

\begin{figure}
\vspace{-3cm}
\hspace{-0.5cm}
\resizebox{10.0cm}{!}{\includegraphics{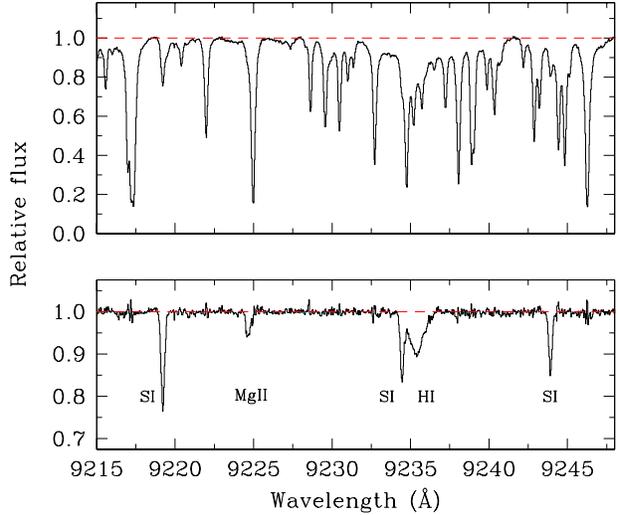}}
\vspace{-2.5cm}
\caption{Portion of the VLT/UVES spectrum of 
\object{HD\,110621} ($V = 9.9$, $\feh = -1.66$) in the far red.
Top panel: the spectrum before removal of the numerous 
telluric \water\ lines which occur in this region.
Bottom panel: the spectrum after 
division by that of the B-type star \object{HR\,5488} using the IRAF 
task {\tt telluric}. The \SI\ triplet, a \MgII\ line and the
Paschen $\zeta$ \HI\ line of \object{HD\,110621} can all be clearly
seen. Note that the spectrum shown has not been corrected
for the radial velocity of the star 
(which introduces a shift of $+6.4$\,\AA ), 
and that the wings of the \HI\ line have been removed 
by the continuum fitting.
Excessive noise in the stellar spectrum is seen
where strong \water\ lines have been removed.}
\label{fig:telluric}
\end{figure}

\section{Observations and data reduction}
\label{observations}
\subsection{UVES observations}
\label{UVES}
Previous determinations of sulphur abundances in metal-poor stars,
including those by Israelian \& Rebolo (\cite{israelian01})
and Takada-Hidai et al. (\cite{takada02}),
have been based on the high excitation ($\chi_{\rm exc} = 7.87$\,eV)
$\lambda \lambda$8694.0, 8694.6 \SI\ lines. 
However, these lines are very weak in metal-poor dwarfs and giants 
(with equivalent widths $W < 3$\,m\AA\ at
$\feh \sim -2.0$), and become vanishingly small at
$\feh \sim -2.5$. In our survey we have taken advantage of the
high efficiency in the far red of the 
VLT/UVES instrument (Dekker et al. \cite{dekker00})
and concentrated on
the stronger \SI\ triplet ($\chi_{\rm exc} = 6.52$\,eV) at 
$\lambda\lambda 9212.9, 9228.1, 9237.5$\,\AA .
The dichroic mode of UVES was used to cover the spectral region
$6700 - 10000$\,\AA\ in the red arm and $3750 - 5000$\,\AA\
in the blue arm; the latter region includes a number of \FeII\
lines suitable for determining the iron abundance, as well as
the \ZnI\ lines at 4722.2 and 4810.5\,\AA .

\begin{figure*}
\vspace{-2.25cm}
\hspace{-0.5cm}
\centerline{\resizebox{16cm}{!}{\includegraphics{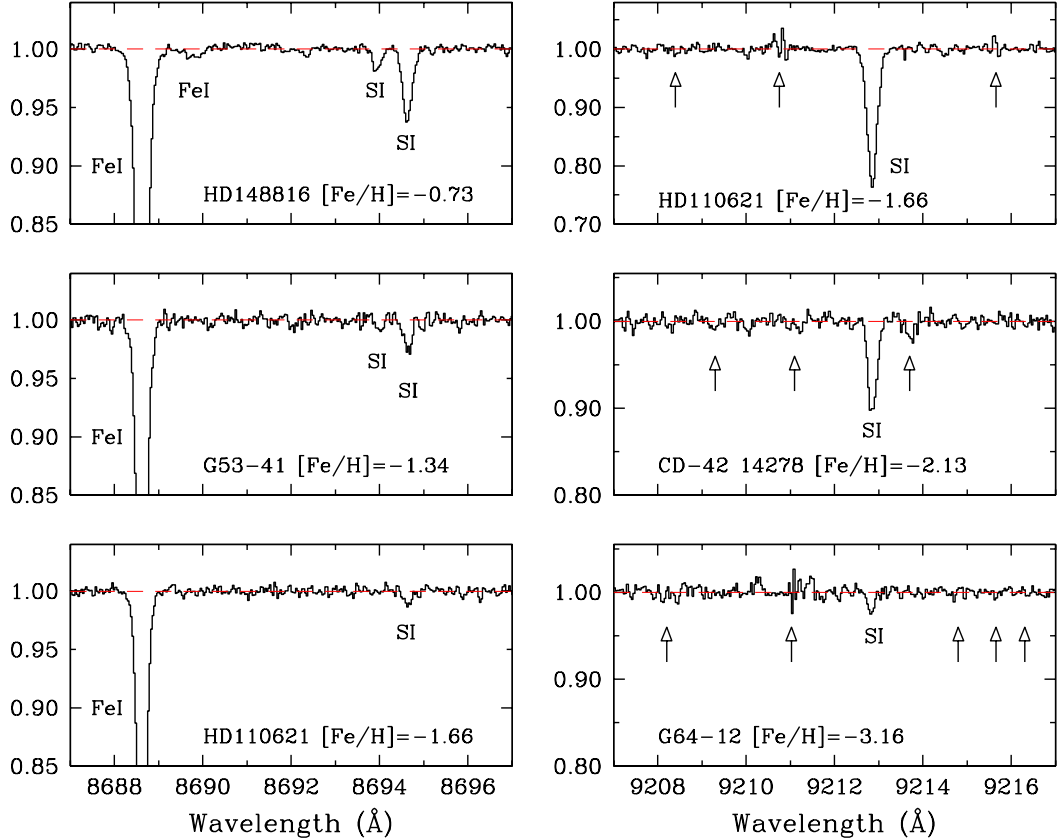}}}
\vspace{-7.0cm}
\caption{Left: A sequence of spectra around the $\lambda \lambda 8694.0, 8694.6$ \SI\ lines.
Right: A sequence of spectra around the $\lambda$9212.9 \SI\ line.
The arrows indicate wavelengths where telluric \water\ lines have
been removed.}
\label{fig:samplespectra}
\end{figure*}

The observations were carried out using the UVES image slicer \#1
(Dekker et al. \cite{dekker02}) which
has an entrance aperture of 2.1 x 2.7 arcsec and
makes 3 slices along the 0.7 arcsec wide entrance slit with a total
length of 8.0 arcsec.
The resulting resolution of the spectra is $\lambda/\Delta\lambda \simeq 60\,000$
with 4 pixels per spectral resolution element. 

Program stars were selected from the Str\"{o}mgren photometric catalogue
of Schuster \& Nissen (\cite{schuster88}) supplemented with a few
very metal-poor stars from Ryan et al. (\cite{ryan99}). The selection
criteria were: $5700 < \teff < 6500$\,K, $3.7 < \logg < 4.4$ and a smooth
distribution of metallicities from [Fe/H]\,=\,$-3.2$ to $-0.6$. 
The UVES spectra were obtained in service mode during the period
March -- June, 2001. The total integration time for a $V = 11$ mag
star was about 60 min, split into three separate exposures so that cosmic 
ray hits could be removed by comparison of the three spectra.
The sky background could be checked adjacent to the stellar spectrum,
but was insignificant (except for emission lines) even for the faintest
star (G64-12, $V = 11.46$). Typical $S/N$ ratios are 300 in the blue spectral
region, 250--300 at the $\lambda \lambda 8694.0, 8694.6$ \SI\ lines,
and 150--200 in the region of the $\lambda\lambda 9212.9, 9228.1, 9237.5$
\SI\ triplet.

\subsection{Reduction of the spectra}
\label{reductions}
The blue spectra were reduced with standard MIDAS routines for
order definition, background subtraction, flat-field correction,
order extraction and wavelength calibration. Bias, dark current and
scattered light corrections are included in the background subtraction.
The spectra were then normalized by a continuum function determined
by fitting a spline curve to a set of pre-selected continuum
windows estimated from the solar atlas. Finally,
correction for the radial velocity of the star, measured from 20--30 narrow
lines, was applied before the measurement of equivalent widths.
For a few stars, it was checked that reduction with NOAO's IRAF
echelle package gave nearly identical results.

The red spectra were reduced in the same way as the blue spectra
but using the IRAF echelle package.
A problem with the 9212--9238\,\AA\ region is the presence
of numerous, strong telluric \water\ lines, see Fig.~\ref{fig:telluric}.
To remove these lines, fast rotating, early-type stars were observed on
each night and reduced in the same way as the program stars.
The IRAF task {\tt telluric} was then applied to remove the telluric lines.
When the early-type star is observed at about the same airmass as the program
star this technique also serves to remove a residual fringing
of a few percent in the red region ($\lambda > 7000$\,\AA )
remaining after the flatfielding of the spectra.

A sequence of spectra of stars  spanning the whole metallicity range
of the sample is shown in Fig.~\ref{fig:samplespectra}. As seen,
the \SI\ line at 9212.9\,\AA\
can be clearly detected in a $\feh \sim -3.2$ star.

\subsection{Equivalent widths}
\label{eqw.widths}
\begin{figure}
\resizebox{\hsize}{!}{\includegraphics{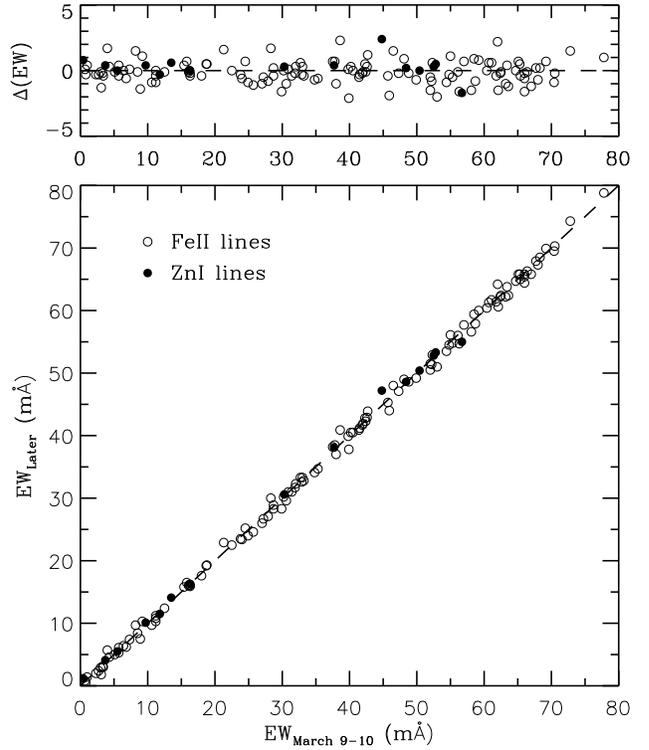}}
\caption{A comparison of equivalent widths of \FeII\
and \ZnI\ lines measured from two sets
of spectra for nine stars observed on
March 9 -- 10 and March 12 -- May 9, 2001, respectively.}
\label{fig:ewblue}
\end{figure}

Equivalent widths of the \FeII , \ZnI\ and \SI\ lines
were measured by Gaussian fitting, or direct integration
if the fit was poor. 
For the present set of data, we could estimate empirically
the accuracy of our equivalent widths 
by comparing two independent sets of measurements for nine stars 
(\object{HD\,103723}, \object{HD\,104004}, \object{HD\,110621},
\object{HD\,121004}, \object{HD\,140283}, \object{HD146296},
\object{G\,16-13}, \object{G\,66-30}, and \object{G64-12}) 
which were observed twice (on different nights).
The first set of observations was carried out on March 9 and 10  
while the second set was made from March 12 to May 9, 2001.
The total exposure time for each star was the same at each epoch.
The comparison of the \FeII\ and \ZnI\ lines is shown in
Fig.~\ref{fig:ewblue}. As can be seen from the figure, 
the data fall close to the 1:1 line
with a standard deviation of only $\pm 0.84$\,m\AA. This corresponds
to an error of $\pm 0.6$\,m\AA\  for one measurement of an equivalent 
width in the blue part of the UVES spectra.

The weak \SI\ lines at 8694.0 and 8694.6\,\AA\ could
be measured in the more metal-rich part of our sample ($\feh > -1.5$).
Among the stronger triplet lines 
$\lambda 9228.1$ falls very close to the center
of the Paschen $\zeta$ \HI\ line 
and its equivalent width could therefore not be measured in a reliable way. The
other two \SI\ lines are, however, ideal for abundance determination in the 
metallicity range $-3 < \feh < -1$.

\begin{figure}
\resizebox{\hsize}{!}{\includegraphics{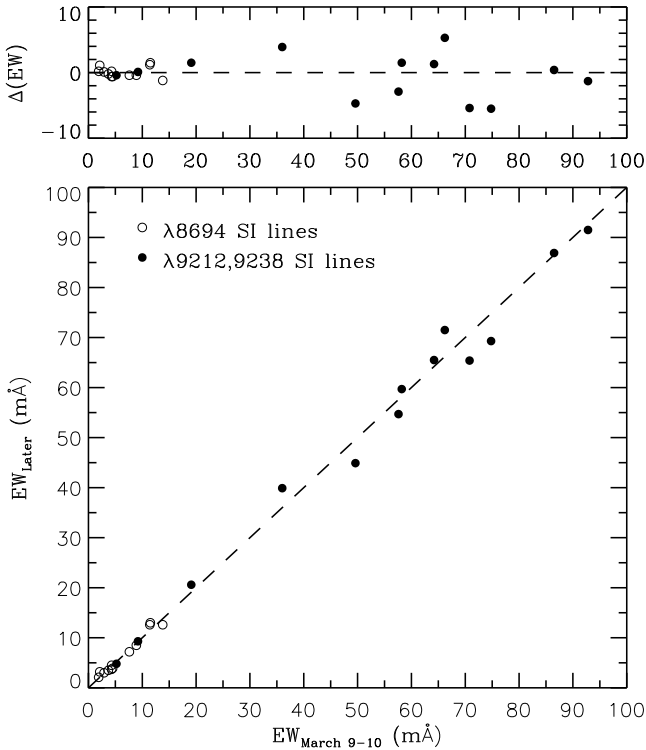}}
\caption{A comparison of equivalent widths of
the $\lambda \lambda 8694.0, 8694.6$ and
$\lambda \lambda 9212.9, 9237.5$\,\AA\ \SI\ lines
measured from two sets of spectra for nine stars.}
\label{fig:ewred}
\end{figure}

The comparison of equivalent widths of \SI\ lines measured from
spectra obtained on different nights is shown in Fig. \ref{fig:ewred}.
Again, the data are evenly distributed around the
1:1 line, but with a larger scatter for the $\lambda \lambda 9212.9, 9237.5$
lines ($\sigma = \pm 3.3$\,m\AA ) than is the case for the 
$\lambda \lambda 8694.0, 8694.6$ pair ($\sigma = \pm 0.8$\,m\AA ). The reason
is the lower S/N in the 
9212 -- 9238\,\AA\ region, made worse by the telluric lines.
Thus, the equivalent widths of
the $\lambda \lambda 9212.9, 9237.5$ lines
have been measured with a precision of $\pm 2.3$\,m\AA\ (one spectrum),
whereas the precision for the weaker $\lambda \lambda 8694.0, 8694$ pair is
as good as $\pm 0.6$\,m\AA .

The error of the measurement of the equivalent widths of the
$\lambda \lambda 9212.9, 9237.5$ \SI\ lines is particular important
for estimating the precision of the sulphur abundance determinations
for the most metal-poor stars, where the lines are weak.
An independent check of the estimated error may be
obtained by comparing the sulphur abundances derived from the
$\lambda 9212.9$ and the $\lambda  9237.5$ \SI\ line, respectively.
For nine stars with $\feh < -2$ having equivalent widths of the
\SI\ lines ranging from 5 to 36\,m\AA\ the mean difference of the
sulphur abundances derived from the two sets of lines is 0.03 dex with a rms
scatter of the difference of $\pm 0.08$\,dex, only. This is, in fact,
a smaller dispersion than expected from the estimated equivalent error
of $\pm 2.3$\,m\AA , which should then be considered as a conservative
estimate.

The measured equivalent widths for 19 \FeII\ lines, the two
\ZnI\ lines and the four \SI\ lines in 34 program stars
are listed in Table A1\footnote{Table A1 is only available in electronic form
at the CDS via anonymous ftp to cdsarc.u-strasbg.fr (130.79.128.5) or via
http://cdsweb.u-strasbg.fr/cgi-bin/qcat?J/A+A/.../...}. 
When no value is given
it is either because the line is too strong or too weak to provide
a reliable abundance, or, in the case of the 
$\lambda \lambda 9212.9, 9237.5$ \SI\ lines, is affected
by residuals from the removal of strong telluric \water\ lines.
In addition to the stars listed in Table A1, three stars,
\object{HD\,99382}, \object{BD\,$-$13\,\,3834} and 
\object{G\,18-54}, were observed,
but they turned out to be double-lined spectroscopic binaries,
and are excluded from our abundance analysis.

\section{Stellar parameters}
\label{parameters}
The determinations of stellar parameters follows the method described by
Nissen et al. (\cite{nissen02}). Here we briefly outline the
procedure and discuss the precision of \teff , \logg\ and
\feh .
The final values of \teff\ and \logg\ are given in Table~\ref{table.abun}  together
with values for \feh\ and the microturbulence as derived from \FeII\ lines
(Sect.~\ref{iron.abun}).
As the calibrations of \teff\ and \logg\ depend somewhat on \feh,
the determination of the parameters is an iterative process.

\subsection{Effective temperature}
\label{teff}
\teff\ was determined from the $b-y$ and $V-K$ colour indices using the
IRFM calibrations of Alonso et al. (\cite{alonso96})
as modified by Nissen et al. (\cite{nissen02}).  
The Str\"omgren $uvby$-$\beta$ photometry was taken from 
Schuster \& Nissen (\cite{schuster88}) for the large majority
of the stars supplemented with unpublished photometry of Schuster et al.
(\cite{schuster03}) for the remaining stars. The $(b-y)_0$ calibration of 
Schuster \& Nissen (\cite{schuster89}) including a zero-point correction
of +0.005\,mag (Nissen \cite{nissen94}) was used to derive the interstellar
reddening excess. If $E(b-y) > 0.015$, the reddening is considered
significant and the $V$ magnitude as well as the $m_1$ and the $c_1$ indices
are corrected according to the relations given by Nissen et al. (\cite{nissen02}).  
These values and the reddening estimates are given in Table~\ref{table.phot}.

The $K$ photometry was taken from Carney (\cite{carney83}),
Alonso et al. (\cite{alonso94}) and the Two Micron
All Sky Survey (2MASS)\footnote{This publication makes use of
data products from the Two Micron All Sky Survey, which is
a joint project of the University of Massachusetts and
the Infrared Processing and Analysis Center/California
Institute of Technology, funded by NASA and the National
Science Foundation.  $K$ magnitudes were taken from
the Second Incremental Release. Additional values from the recent full
release have only a marginal effect on the derived \teff\ and
abundances.}.
In this connection we note that the 2MASS $K$ magnitudes are on the so-called
$K$-short system (Cutri et al. \cite{cutri03}), whereas those of Carney
(\cite{carney83}) are on the CIT (California Institute of Technology)
system and Alonso et al.'s (\cite{alonso94}) magnitudes are on the TCS
(Telescope Carlos S\'{a}nchez) system. Small differences between
these systems may exist. Carpenter (\cite{carpenter01})
has derived an average transformation $K$(2MASS) = $K$(CIT) $-$ 0.024
without any colour term.  For our actual sample of metal-poor turnoff
and subgiant stars the agreement is even better. Twelve stars with both
CIT and 2MASS photometry have a mean difference in $K$ of 0.002 only
with a rms-dispersion of the difference of 0.026. For 16 stars with
both Alonso et al. and 2MASS photometry the mean $K$ difference
(TCS - 2MASS) is $-0.019$ with a dispersion of 0.030. Hence, the three
systems agree within $\pm 0.02$\,mag in $K$ for our sample, which according
to the Alonso et al. (\cite{alonso96}) \teff\
calibration of $V-K$ corresponds to an error of $\pm 20$\,K in \teff .
The corresponding effect on the derived abundances is small
compared to other error sources as seen from Table~\ref{table.change}.
We have, therefore,
adopted the straight mean of the $K$ magnitudes if two or three
sources were available. Table~\ref{table.phot} lists the corresponding
$V-K$ value corrected for interstellar reddening
according to the relation $E(V-K) = 2.7 E(B-V) = 3.8 E(b-y)$
(Savage \& Mathis \cite{savage79}), if $E(b-y) > 0.015$.

Comparing the \teff\ values determined from $b-y$ and $V-K$
we find a mean difference ($\Delta = \teff (b-y) - \teff (V-K)$)
$<\Delta> = +44$\,K for 25 stars with a standard deviation of 
$\sigma (\Delta) = 78$\,K (one star). Similar values were 
found for the sample of metal-poor stars in Nissen et al. (\cite{nissen02}).
The mean value of $\teff (b-y)$ and $\teff (V-K)$ is adopted for 
the 25 stars. For nine stars without $K$ photometry, a value
$\teff (b-y) - 22$~K 
has been adopted. The typical observational errors are 0.007 mag in 
$b-y$ and 0.05 mag in $V-K$, which correspond to an error 
of $\pm 50$~K in \teff\ in either case. Taking into account the 
uncertainty in the reddening estimate (also corresponding to about
$\pm 50$~K in \teff ), we estimate the 1-$\sigma$
statistical error of $\teff$ to be around 70~K.

\begin{table*}
\caption[ ]{Str\"{o}mgren photometry, colour excess, $V-K$ index,
\teff\ from $b-y$ and $V-K$, spectroscopic value of \feh , and
absolute magnitudes derived from the Str\"{o}mgren photometry and
from the Hipparcos parallax including an error corresponding to the
parallax error. If
$E(b-y) > 0.015$, the $V$ magnitudes and the photometric indices have been
corrected for interstellar absorption.}
\label{table.phot}
\setlength{\tabcolsep}{0.12cm}
\begin{tabular}{lrccccrccccccc}
\noalign{\smallskip}
\hline
\noalign{\smallskip}
Star & $V_0$ & $(b-y)_0$ & $m_0$ & $c_0$ & $\beta$ & $E_{b-y}$ & $(V-K)_0$ & $T_{e,b-y}$ & $T_{e,V-K}$ &
 \feh & \mvphot  & \mvpar & $\pm \sigma$ \\
     & & & & & & & & [K] & [K] & & & &  \\
\noalign{\smallskip}
\hline
\noalign{\smallskip}
BD$-13\degr3442 $&  10.153 & 0.277 & 0.059 & 0.379 & 2.622 &  0.031 &  1.129 &  6526 &  6475 & $-$2.61 &  4.03 &       &      \\
CD$-30\degr18140$&   9.859 & 0.302 & 0.053 & 0.340 & 2.606 &  0.021 &  1.214 &  6285 &  6259 & $-$1.88 &  3.93 &  4.18 & 0.46 \\
CD$-35\degr14849$&  10.568 & 0.321 & 0.040 & 0.293 & 2.603 &  0.010 &        &  6147 &       & $-$2.41 &  4.24 &       &      \\
CD$-42\degr14278$&  10.216 & 0.361 & 0.040 & 0.229 & 2.576 &  0.009 &        &  5834 &       & $-$2.12 &  4.83 &       &      \\
HD\,103723   &       9.947 & 0.329 & 0.104 & 0.306 & 2.625 &  0.027 &        &  6064 &       & $-$0.82 &  4.37 &  4.36 & 0.46 \\
HD\,105004   &      10.194 & 0.360 & 0.124 & 0.261 & 2.616 &  0.027 &  1.337 &  5830 &  6007 & $-$0.86 &  4.78 &       &      \\
HD\,106038   &      10.179 & 0.342 & 0.092 & 0.264 & 2.583 & $-$0.018 &  1.408 &  5940 &  5898 & $-$1.42 &  4.69 &  4.99 & 0.36 \\
HD\,108177   &       9.671 & 0.330 & 0.059 & 0.287 & 2.594 & $-$0.001 &  1.342 &  6047 &  6033 & $-1$.74 &  4.36 &  4.87 & 0.26 \\
HD\,110621   &       9.906 & 0.337 & 0.067 & 0.324 & 2.595 &  0.008 &        &  6011 &       & $-$1.66 &  3.75 &  4.15 & 0.46 \\
HD\,121004   &       9.031 & 0.395 & 0.140 & 0.268 & 2.588 &  0.012 &        &  5617 &       & $-$0.77 &  4.89 &  5.15 & 0.18 \\
HD\,140283   &       7.213 & 0.380 & 0.033 & 0.284 & 2.564 &  0.015 &  1.580 &  5734 &  5646 & $-$2.42 &  3.53 &  3.42 & 0.12 \\
HD\,146296   &       9.761 & 0.383 & 0.143 & 0.268 & 2.581 & $-$0.004 &        &  5693 &       & $-$0.74 &  4.87 &  4.27 & 0.38 \\
HD\,148816   &       7.287 & 0.366 & 0.117 & 0.312 & 2.582 & $-$0.005 &  1.448 &  5822 &  5824 & $-$0.73 &  4.20 &  4.22 & 0.08 \\
HD\,160617   &       8.733 & 0.347 & 0.051 & 0.331 & 2.584 &  0.011 &  1.408 &  5953 &  5909 & $-$1.79 &  3.27 &  3.42 & 0.31 \\
HD\,179626   &       9.210 & 0.373 & 0.095 & 0.293 & 2.588 &  0.015 &  1.560 &  5750 &  5649 & $-$1.14 &  4.24 &  3.59 & 0.39 \\
HD\,181743   &       9.687 & 0.351 & 0.052 & 0.224 & 2.582 &  0.000 &        &  5885 &       & $-$1.93 &  4.97 &  4.95 & 0.34 \\
HD\,188031   &      10.148 & 0.328 & 0.058 & 0.310 & 2.592 & $-$0.001 &        &  6076 &       & $-$1.79 &  4.02 &       &      \\
HD\,193901   &       8.660 & 0.383 & 0.099 & 0.221 & 2.568 & $-$0.002 &  1.533 &  5652 &  5691 & $-$1.12 &  5.16 &  5.46 & 0.12 \\
HD\,194598   &       8.354 & 0.344 & 0.091 & 0.269 & 2.588 & $-$0.011 &        &  5928 &       & $-$1.17 &  4.63 &  4.62 & 0.15 \\
HD\,215801   &      10.044 & 0.334 & 0.049 & 0.330 & 2.572 & $-$0.018 &  1.402 &  6071 &  5940 & $-$2.29 &  3.47 &       &      \\
LP\,815$-$43 &      10.774 & 0.272 & 0.054 & 0.376 & 2.623 &  0.032 &  1.117 &  6558 &  6508 & $-$2.67 &  4.17 &       &      \\
G\,011$-$044 &      11.091 & 0.335 & 0.061 & 0.248 & 2.588 & $-$0.009 &  1.381 &  6023 &  5967 & $-$2.09 &  4.80 &       &      \\
G\,013$-$009 &       9.916 & 0.292 & 0.054 & 0.369 & 2.609 &  0.019 &  1.195 &  6402 &  6319 & $-$2.27 &  3.68 &        &      \\
G\,016$-$013 &       9.951 & 0.387 & 0.129 & 0.286 & 2.583 &  0.008 &  1.627 &  5676 &  5529 & $-$0.76 &  4.60 &       &      \\
G\,018$-$039 &      10.392 & 0.346 & 0.073 & 0.286 & 2.581 & $-$0.007 &  1.412 &  5926 &  5894 & $-$1.52 &  4.32 &       &      \\
G\,020$-$008 &       9.948 & 0.356 & 0.047 & 0.251 & 2.574 &  0.001 &  1.465 &  5879 &  5831 & $-$2.28 &  4.58 &  4.56 & 0.43 \\
G\,024$-$003 &      10.393 & 0.346 & 0.062 & 0.268 & 2.585 &  0.017 &  1.414 &  5925 &  5894 & $-$1.67 &  4.51 &       &      \\
G\,029$-$023 &      10.230 & 0.339 & 0.059 & 0.332 & 2.590 &  0.008 &  1.400 &  6010 &  5923 & $-$1.80 &  3.49 &       &      \\
G\,053$-$041 &      11.022 & 0.356 & 0.083 & 0.271 & 2.589 &  0.004 &  1.460 &  5848 &  5810 & $-$1.34 &  4.54 &       &      \\
G\,064$-$012 &      11.330 & 0.282 & 0.046 & 0.330 & 2.622 &  0.030 &  1.114 &  6458 &  6563 & $-$3.17 &  4.54 &       &      \\
G\,064$-$037 &      11.144 & 0.299 & 0.057 & 0.329 & 2.626 &  0.009 &  1.235 &  6328 &  6308 & $-$3.12 &  4.21 &       &      \\
G\,066$-$030 &      10.950 & 0.287 & 0.077 & 0.354 & 2.634 &  0.018 &  1.177 &  6381 &  6311 & $-$1.52 &  4.25 &       &      \\
G\,126$-$062 &       9.478 & 0.330 & 0.063 & 0.327 & 2.588 & $-$0.004 &  1.455 &  6060 &  5825 & $-$1.64 &  3.75 &  4.11 & 0.37 \\
G\,186$-$026 &      10.829 & 0.306 & 0.041 & 0.339 & 2.608 &  0.011 &  1.229 &  6280 &  6266 & $-$2.62 &  3.73 &  5.20  & 0.50 \\
\hline
\end{tabular}
\end{table*}

\subsection{Surface gravity}
\label{logg}
The surface gravity was determined from the fundamental relation
\begin{eqnarray}
 \log \frac{g}{g_{\sun}} & = & \log \frac{\cal{M}}{\cal{M}_{\sun}}
 + 4 \log \frac{\teff}{T_{\rm eff,\sun}} + \\
    &   & 0.4 (M_{bol} - M_{bol,\sun})  \nonumber
\end{eqnarray}
where $\cal{M}$ is the mass of the star and $M_{bol}$ the absolute bolometric
magnitude.

The absolute visual magnitude \Mv\ was determined from a new calibration
of the Str\"{o}mgren indices derived by Schuster et al.
(\cite{schuster03}) on the basis of Hipparcos parallaxes, and
also directly from the Hipparcos parallax (ESA \cite{esa97})
if available with an error $\sigma (\pi) / \pi < 0.3$. Column 12 and 13 of
Table~\ref{table.phot}  list the photometric and the parallax based values of \Mv .
They compare reasonably well for the 18 stars 
having \mvpar . Excluding one star, G\,186-26, with a large difference
but also with the largest uncertainty of \mvpar , the mean difference
($\Delta = \mvphot - \mvpar$) is $<\!\Delta\!> = -0.07$\,mag
with a standard deviation of $\sigma (\Delta) = 0.32$\,mag.
The corresponding error in \Mv\ is $\pm 0.22$\,mag. This induces
an error of about $\pm 0.1$\,dex in \logg . Taking into account possible
errors in the bolometric correction (adopted from Alonso et al. (\cite{alonso95}))
and the mass as derived by interpolating in the \Mv -- \logteff\
diagram between the $\alpha$-element enhanced evolutionary tracks of
VandenBerg et al. (\cite{vandenberg00}), we estimate that the error of
\logg\ is about $\pm 0.15$\,dex.

\section{Abundances}
\label{abundances}
The determination of abundances is based on $\alpha$-element enhanced
(\alphafe = +0.4, $\alpha$ = O, Ne, Mg, Si, S, Ca, and Ti) 
1D model atmospheres with \teff , \logg\ and \feh\ values
as given in Table~\ref{table.abun} and a microturbulence of 1.5 \kmprs .
The models were computed with the MARCS code using updated continuous
opacities (Asplund et al. \cite{asplund97}) and including UV line blanketing
by millions of absorption lines.
LTE is assumed both in constructing the models and in deriving abundances.

The Uppsala abundance analysis program, EQWIDTH,
was used to calculate theoretical equivalent widths
from the models. An elemental
abundance is determined by requiring that the calculated
equivalent width should match the observed one.
When more than one line of the same species were
measured for a star, the mean value is adopted by
giving equal weight to each line.

\subsection{Iron}
\label{iron.abun}
By inspecting the blue spectra, 19 apparently unblended \FeII\ lines
were selected for determining the iron abundance. 
Differential $\log gf$ values of these lines were derived by an inverted abundance
analysis of four  `standard' stars, \object{HD\,103723},
\object{HD\,160617}, \object{G\,13-09} and \object{HD\,140283}
adopting iron abundances of \feh = $-0.79$, $-1.79$, $-2.30$
and $-2.42$, respectively, from Nissen et al. (\cite{nissen02}).
In this calculation, only lines with equivalent widths between
5 and 50\,m\AA\ were included. If a differential $\log gf$ value
is available for more than one of the four standard stars, then
the mean value (given in Table~\ref{table.lines}) is adopted.
We note that the \feh\ values of Nissen et al. (\cite{nissen02})
are based on very weak \FeII\ lines 
in the red spectral region measured from extremely high S/N spectra
and analyzed differentially to the solar flux spectrum. The same procedure
could not be applied to the blue \FeII\ lines because they are quite strong
in the solar spectrum and often blended with weak lines
making the solar equivalent width measurement difficult.

In calculating abundances from the \FeII\ lines we adopted the
Uns\"{o}ld (\cite{unsold55}) approximation to the Van der Waals
interaction constant with an enhancement factor $E_{\gamma} = 2.5$.
But the adopted value of $E_{\gamma}$ does not have a large
effect on \feh\ since most \FeII\ lines in the stars
are weak, especially in metal-poor stars. For the most metal-rich
stars in our sample the change of \feh\ is  around +0.03\,dex
when $E_{\gamma}$ is changed from 2.5 to 1.5.

In the more metal-poor stars ($\feh < -1.5$) the \FeII\ lines are so
weak that the derived metallicity is practically independent of
the microturbulence. For such stars we have assumed
$\xi_{\rm micro} = 1.5$ \kmprs . For the more metal-rich stars
$\xi_{\rm micro}$ has been determined by requesting that the derived \feh\ values
should be independent of equivalent width.
The error of $\xi_{\rm micro}$ is about $\pm 0.2$~\kmprs .

The present sample has 10 stars in common with 
the sample of Nissen et al. (\cite{nissen02}).
The metallicities of these stars range from $\feh = -0.75$ to $-2.70$.
The mean difference (present $-$ 2002) is
$<\Delta \feh> = -0.01$ with a standard deviation of $\pm 0.04$\,dex. 
This small scatter shows that precise \feh\ values have been determined
in both papers.

\begin{table}
\caption[ ]{The list of lines used to determine the abundances of Fe, S and Zn.
Measured equivalent widths are given for three representative stars:
(a) \object{HD\,194598}, $\feh = -1.17$, 
(b) \object{HD\,160617}, $\feh = -1.79$, and 
(c) \object{HD\,140283}, $\feh = -2.42$.}
\label{table.lines}
\setlength{\tabcolsep}{0.12cm}
\begin{tabular}{cccrrrr}
\noalign{\smallskip}
\hline
\noalign{\smallskip}
 ID & Wavelength & Exc.\,Pot. & $\log gf$ & $W$(a) &  $W$(b) &  $W$(c) \\
    & \AA        &  eV        &           & m\AA   & m\AA    & m\AA   \\
\hline
\noalign{\smallskip}
\FeII & 4128.74&2.58& $-$3.73& 15.0&  5.0&      \\
\FeII & 4178.86&2.58& $-$2.61& 55.8& 39.7& 17.8 \\
\FeII & 4233.16&2.58& $-$2.01& 81.3& 68.1& 41.8 \\
\FeII & 4416.83&2.78& $-$2.65& 46.3& 29.5& 11.2 \\
\FeII & 4489.18&2.83& $-$2.96& 32.5& 17.3&  5.5 \\
\FeII & 4491.41&2.85& $-$2.80& 37.9& 21.4&  7.4 \\
\FeII & 4508.29&2.85& $-$2.41& 55.1& 37.4& 16.2 \\
\FeII & 4515.34&2.84& $-$2.56& 48.9& 31.2& 12.4 \\
\FeII & 4520.23&2.81& $-$2.66& 46.0& 28.9& 11.0 \\
\FeII & 4522.63&2.84& $-$2.22& 64.6& 48.3& 22.5 \\
\FeII & 4541.52&2.85& $-$3.04& 28.4& 14.4&  4.5 \\
\FeII & 4555.89&2.83& $-$2.43& 53.9& 37.5& 16.1 \\
\FeII & 4576.34&2.84& $-$3.01& 28.5& 15.0&  5.0 \\
\FeII & 4582.83&2.84& $-$3.24& 20.9&  9.8&  3.2 \\
\FeII & 4583.83&2.81& $-$1.91& 82.1& 63.8& 37.9 \\
\FeII & 4620.52&2.83& $-$3.36& 17.5&  7.4&  2.5 \\
\FeII & 4656.98&2.89& $-$3.66&  8.1&  3.1&      \\
\FeII & 4666.75&2.83& $-$3.31& 17.9&  7.5&  2.9 \\
\FeII & 4923.93&2.89& $-$1.45&     & 82.9& 58.5 \\
\ZnI & 4722.16&4.03& $-$0.39& 23.8&  8.6&  3.9 \\
\ZnI & 4810.54&4.08& $-$0.17& 29.0& 13.5&  5.5 \\
\SI   & 8693.96&7.87& $-$0.56&  1.9&     &      \\
\SI   & 8694.64&7.87&    0.03&  5.1&  4.1&      \\
\SI   & 9212.87&6.52&    0.38& 77.0& 57.4& 19.8 \\
\SI   & 9237.54&6.52&    0.01& 53.2& 34.4&  9.2 \\
\noalign{\smallskip}
\hline
\end{tabular}
\end{table}

\subsection{Sulphur}
\label{sulphur.abun}
The $\log gf$ values of the \SI\ lines listed in Table~\ref{table.lines}
are adopted from the Coulomb approximation calculations of
Lambert \& Luck (\cite{lambert78}). From the equivalent widths
of the $\lambda \lambda$8694.0, 8694.6 \SI\ lines in the solar flux
spectrum (Kurucz et. al. \cite{kurucz84}) of 11.0 and 28.8\,m\AA ,
respectively, we derive a solar sulphur abundance of log$\epsilon$(S) = 7.20.
This is also the meteoritic abundance (Grevesse \& Sauval \cite{grevesse98}),
but given possible errors in the $\log gf$ values (the NIST database lists
values that are 0.04 dex higher than those of Lambert \& Luck)
the exact agreement is quite fortuitous. Anyhow, when
deriving the differential sulphur abundance with respect to
the Sun, i.e. \sh , a possible error in $\log gf$ is of no concern.
Furthermore, since the $\lambda \lambda 8694.0, 8694.6$ \SI\ lines are quite 
weak and apparently without any blends in the solar spectrum,
the derived \sh\ is not subject to the kind of systematic errors 
which may affect \znh\ as discussed in Sect~\ref{zinc.abun}. 

\begin{figure}
\resizebox{\hsize}{!}{\includegraphics{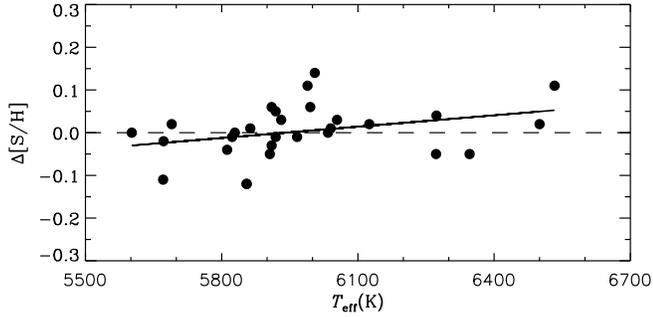}}
\caption{The difference in S abundances derived from
the $\lambda 9212.9$ and  $\lambda 9237.5$ \SI\ lines.
The least squares fit to the data is shown. The rms deviation
between the two sets of abundances is $\pm 0.06$\,dex.}
\label{fig:S.9212-9237}
\end{figure}

The flux around the $\lambda 9237.5$ \SI\ line is slightly
depressed relative to the true continuum by the broad wing
of the Paschen $\zeta$ \HI\ line at
9229\,\AA . In this connection, we note that
the equivalent widths of the \SI\ lines were measured relative to the
local, apparent continuum around the lines. Furthermore, when deriving
the sulphur abundance we neglected the contribution of the \HI\ line
to the line absorption coefficient. For a few representative stars
a spectrum synthesis of the region
around the $\lambda 9237.5$ \SI\ line was carried out including the line 
absorption contribution from the Paschen $\zeta$ \HI\ line 
calculated as described by Seaton (\cite{seaton90}). This exercise shows that the
correction to the sulphur abundance derived from the equivalent
width of the $\lambda 9237.5$ \SI\ line is at most about +0.07~dex
for our hottest stars and decreases with decreasing \teff . In the case of
the $\lambda 9212.9$ \SI\ line, which is further away from the \HI\
line, the corresponding correction is negligible. An empirical
confirmation of the effect is seen in Fig.~\ref{fig:S.9212-9237},
where the difference of sulphur abundances derived from the $\lambda 9212.9$
and $\lambda 9237.5$ lines is plotted as a function of \teff .
Given the small size of the effect and the uncertainty in the
line broadening theory of this high-series Paschen \HI\ line, 
we have not included any correction for the \HI\ line absorption. In this
connection, we note that for the two most metal-poor stars in
our sample, \object{G\,64-12} and \object{G\,64-37},
only the $\lambda 9212.9$ \SI\ line
could be detected and used for deriving the S abundance.

The $\lambda \lambda 9212.9, 9237.5$ \SI\ lines are too
strong in the solar spectrum and the spectral region 
is too crowded with telluric \water\ lines to allow a reliable
solar S abundance to be determined from these lines.
In order to check that this pair 
provides S abundances on the same scale as the $\lambda \lambda 8694.0, 8694.6$
lines, we make use of the fact that for 18 of the 34 program stars
the S abundance has been derived from both the 
$\lambda \lambda 8694.0, 8694.6$ and the  $\lambda \lambda 9212.9, 9237.5$
pairs. The mean abundance difference is 0.03 dex and the standard
deviation is $\pm 0.08$\,dex. Hence, we are confident that the two
sets of \SI\ lines provide \sh\ values on the same scale.

\subsection{Zinc}
\label{zinc.abun}
The zinc abundance is based on two \ZnI\ lines at 4722.2
and 4810.5\,\AA . The $\log gf$ values ($-$0.39 and $-$0.17 respectively)
were adopted from the calculations of
Bi\'{e}mont \& Godefroid (\cite{biemont80}).
From equivalent widths measured in the solar flux
spectrum (Kurucz et. al. \cite{kurucz84}) of 70.3 and 78.3\,m\AA ,
respectively, and using a solar model atmosphere computed with the
same code as that for the stars, we derive a solar zinc abundance
of $\log \epsilon$(Zn) = 4.57. This is 0.10 dex lower than the
meteoritic abundance of 4.67 $\pm 0.04$
(Grevesse \& Sauval \cite{grevesse98}). The same problem was
encountered by Bi\'{e}mont \& Godefroid (\cite{biemont80}),
who derived a solar
photospheric zinc abundance of log$\epsilon$(Zn) = 4.60 from
six \ZnI\ lines (including $\lambda 4722.2$ and $\lambda 4810.5$)
using the Holweger-M\"{u}ller (1974) model of the Sun. This difference
between photospheric and meteoritic Zn abundances may well
be due to a systematic error in the $\log gf$ values of the \ZnI\
lines. However, by determining differential Zn abundances with respect
to the Sun, i.e. \znh , using
our solar photospheric abundance of $\log \epsilon$(Zn) = 4.57
as a reference, errors in $\log gf$ will cancel.
On the other hand, it should be noted that the \ZnI\ lines in the solar
spectrum are quite strong and hence subject to errors in the adopted
solar flux microturbulence (1.15\,\kmprs) and the Van der Waals damping
constant (an enhancement factor $E_{\gamma} = 1.5$ relative to
the Uns\"{o}ld approximation was adopted). The solar equivalent
widths of the \ZnI\ $\lambda \lambda 4722.2, 4810.5$ lines are also
quite uncertain due to blends in their wings. Hence, the scale
of the \znh\ values for our metal-poor stars could be in error
by up to 0.10 dex.

\subsection{Results}
\label{results}
The abundances \feh , \sh\ and \znh\ derived for
the stars in our sample are listed in 
Table~\ref{table.abun} together with the stellar
atmospheric parameters.
For the most metal-poor stars the \ZnI\
lines are too weak ($W < 2$\,m\AA ) to provide reliable Zn
abundances, and the value of \znh\ is not given.

\begin{table}
\caption[ ]{ The derived values of the effective temperature,
surface gravity, microturbulence, and the abundances of iron, sulphur and zinc.}
\label{table.abun}
\setlength{\tabcolsep}{0.12cm}
\begin{tabular}{lccccccccc}
\noalign{\smallskip}
\hline
\noalign{\smallskip}
Star & \teff & \logg & $\xi_{\rm turb}$ &  \feh & \sh & \znh \\
     & [K]     & [cgs]   & [km/s] &       &  &      \\
\noalign{\smallskip}
\hline
\noalign{\smallskip}
\object{BD$-13\degr3442 $}&    6500 &   4.16 &   1.5 &    $-$2.61 & $-2.31$ &         \\
\object{CD$-30\degr18140$}&    6272 &   4.13 &   1.5 &    $-$1.88 & $-1.63$ & $-1.82$ \\
\object{CD$-35\degr14849$}&    6125 &   4.11 &   1.5 &    $-$2.41 & $-2.10$ & $-2.27$ \\
\object{CD$-42\degr14278$}&    5812 &   4.25 &   1.5 &    $-$2.12 & $-1.82$ & $-2.04$ \\
\object{HD\,103723}   &        6040 &   4.26 &   1.3 &    $-$0.82 & $-0.84$ & $-0.95$ \\
\object{HD\,105004}   &        5919 &   4.36 &   1.2 &    $-$0.86 & $-0.85$ & $-0.86$ \\
\object{HD\,106038}   &        5919 &   4.30 &   1.2 &    $-$1.42 & $-1.00$ & $-1.25$ \\
\object{HD\,108177}   &        6034 &   4.25 &   1.5 &    $-$1.74 & $-1.39$ & $-1.65$ \\
\object{HD\,110621}   &        5989 &   3.99 &   1.5 &    $-$1.66 & $-1.34$ & $-1.63$ \\
\object{HD\,121004}   &        5595 &   4.31 &   1.0 &    $-$0.77 & $-0.43$ & $-0.58$ \\
\object{HD\,140283}   &        5690 &   3.69 &   1.5 &    $-$2.42 & $-2.11$ & $-2.35$ \\
\object{HD\,146296}   &        5671 &   4.17 &   1.2 &    $-$0.74 & $-0.54$ & $-0.69$ \\
\object{HD\,148816}   &        5823 &   4.14 &   1.2 &    $-$0.73 & $-0.49$ & $-0.61$ \\
\object{HD\,160617}   &        5931 &   3.77 &   1.5 &    $-$1.79 & $-1.40$ & $-1.81$ \\
\object{HD\,179626}   &        5699 &   3.92 &   1.2 &    $-$1.14 & $-0.85$ & $-1.14$ \\
\object{HD\,181743}   &        5863 &   4.32 &   1.5 &    $-$1.93 & $-1.59$ & $-1.94$ \\
\object{HD\,188031}   &        6054 &   4.03 &   1.5 &    $-$1.79 & $-1.47$ & $-1.76$ \\
\object{HD\,193901}   &        5672 &   4.38 &   1.0 &    $-$1.12 & $-0.95$ & $-1.18$ \\
\object{HD\,194598}   &        5906 &   4.25 &   1.3 &    $-$1.17 & $-0.99$ & $-1.22$ \\
\object{HD\,215801}   &        6005 &   3.81 &   1.5 &    $-$2.29 & $-2.01$ & $-2.21$ \\
\object{LP\,815$-$43} &        6533 &   4.25 &   1.5 &    $-$2.67 & $-2.43$ &         \\
\object{G\,011$-$044} &        5995 &   4.29 &   1.5 &    $-$2.09 & $-1.71$ & $-1.90$ \\
\object{G\,013$-$009} &        6360 &   4.01 &   1.5 &    $-$2.27 & $-1.81$ & $-2.20$ \\
\object{G\,016$-$013} &        5602 &   4.17 &   1.0 &    $-$0.76 & $-0.47$ & $-0.81$ \\
\object{G\,018$-$039} &        5910 &   4.09 &   1.5 &    $-$1.52 & $-1.15$ & $-1.44$ \\
\object{G\,020$-$008} &        5855 &   4.16 &   1.5 &    $-$2.28 & $-1.85$ & $-2.15$ \\
\object{G\,024$-$003} &        5910 &   4.16 &   1.5 &    $-$1.67 & $-1.29$ & $-1.61$ \\
\object{G\,029$-$023} &        5966 &   3.82 &   1.5 &    $-$1.80 & $-1.50$ & $-1.83$ \\
\object{G\,053$-$041} &        5829 &   4.15 &   1.3 &    $-$1.34 & $-1.05$ & $-1.29$ \\
\object{G\,064$-$012} &        6511 &   4.39 &   1.5 &    $-$3.17 & $-2.81$ &         \\
\object{G\,064$-$037} &        6318 &   4.16 &   1.5 &    $-$3.12 & $-2.78$ &         \\
\object{G\,066$-$030} &        6346 &   4.24 &   1.5 &    $-$1.52 & $-1.28$ & $-1.52$ \\
\object{G\,126$-$062} &        5943 &   3.97 &   1.5 &    $-$1.64 & $-1.31$ & $-1.71$ \\
\object{G\,186$-$026} &        6273 &   4.25 &   1.5 &    $-$2.62 & $-2.43$ &         \\
\hline
\end{tabular}
\end{table}

In Fig.~\ref{fig:SFe}, \sfe\ is plotted as a function of \feh .
All our program stars have space velocities that are typical of
halo stars, and are shown with filled circles. In addition, we have
plotted disk stars from Chen et al. (\cite{chen02}), who determined
sulphur abundances from five weak \SI\ lines including the  
$\lambda \lambda 8694.0, 8694.6$ pair.

Fig.~\ref{fig:ZnFe} shows \znfe\ vs. \feh . Here the data
for the disk stars are from Chen et al. (\cite{chen03}),
who observed the weak \ZnI\ line ($\chi_{\rm exc} = 5.79$\,eV)
at 6362.35\,\AA\ and determined \znh\ by a differential
model atmosphere analysis 
with respect to the Sun. This Zn line lays in the midst of a very
broad and weak \CaI\  auto-ionization line 
(Mitchell \& Mohler \cite{mitchell65}), but its equivalent width can 
still be measured reliably with respect to the local continuum.
In the solar flux spectrum (Kurucz et al. \cite{kurucz84})
its equivalent width is close to 22.0\,m\AA\ and the line appears 
unblended. Hence this line is highly suitable for determining
Zn abundances in disk stars. The 
$\lambda 6362.35$ line is not covered by the UVES spectra
of the present program, but in the very high S/N spectra of
Nissen et al. (\cite{nissen02}) we were able to detect the line for
three of the more metal-rich halo
stars and to derive \znh\ in a differential
analysis with respect to the Sun: 
\object{HD\,103723}, $W = 4.9$m\AA , $\znh = -0.85$;
\object{HD\,106038}, $W = 2.6$m\AA , $\znh = -1.18$; and
\object{HD\,121004}, $W = 9.5$m\AA , $\znh = -0.53$.
These Zn abundances compare reasonably well with those given in
Table~\ref{table.abun} (the differences being $\leq 0.1$\,dex),
indicating that the values of \znh\ derived in the 
halo stars from the \ZnI\ $\lambda \lambda 4722.2, 4810.5$
lines are approximately on the same scale as \znh\ for the disk stars.

Finally, Fig.~\ref{fig:SZn} shows \szn\ vs. \znh . Here the disk
stars are those from Chen et al. (\cite{chen02}, \cite{chen03})
for which both S and Zn abundances are available.

\subsection{Statistical abundance errors}
\label{stat.errors}
The main contributions to the random errors in the abundances arise
from the errors in the equivalent width measurements and the
uncertainties in the model atmosphere parameters. The level of
the latter contribution has been estimated by changing 
\teff\ of the stellar models by 100\,K, \logg\ by 0.2\,dex and
the metallicity by 0.2\,dex.
Table~\ref{table.change} shows the corresponding changes in
the various abundance ratios. Evidently, $\Delta \teff = 100$\,K
has a negligible effect of \feh , and rather small effects
on the other abundance ratios, except \szn .
\feh\ and \sh\ are quite sensitive to \logg\ but
the \logg\ effects nearly cancel in the S/Fe ratio.
We also see that the effect of changing the metal content
of the model atmosphere is negligible.
Finally, we note that the possible error in the microturbulence
parameter only has a  significant effect on the derived abundances for
the more metal-rich stars.

\begin{figure}
\resizebox{\hsize}{!}{\includegraphics{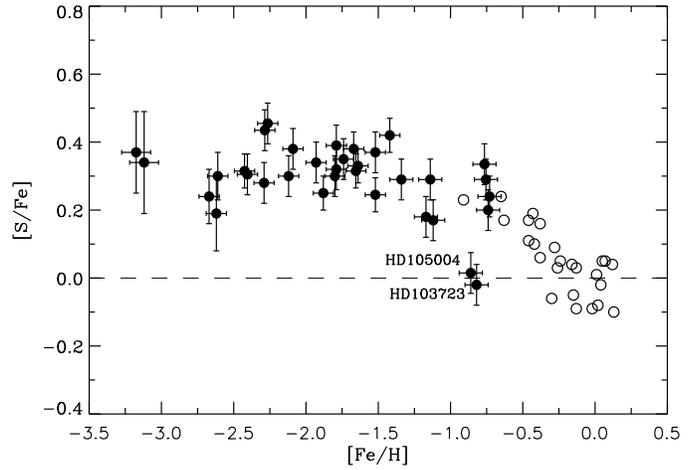}}
\caption{\sfe\ vs. \feh . Filled circles refer to
halo stars from the present program. Open circles are
disk stars with abundances from Chen et al. (\cite{chen02}).
The error bars indicate 1-$\sigma$ statistical errors of
\sfe\ and \feh .}
\label{fig:SFe}
\end{figure}

\begin{figure}
\resizebox{\hsize}{!}{\includegraphics{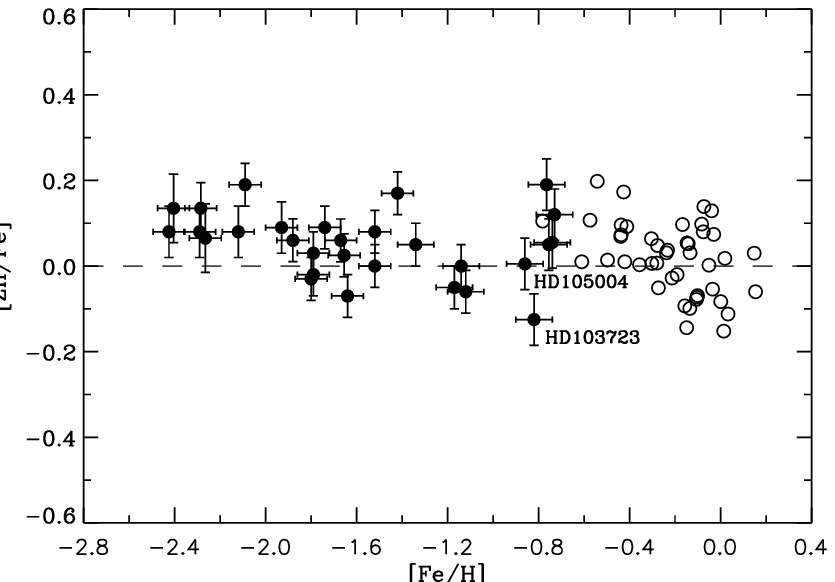}}
\caption{\znfe\ vs. \feh . Filled circles: halo stars;
open circles: disk stars from Chen et al. (\cite{chen03}).
The error bars indicate 1-$\sigma$ statistical errors of
\znfe\ and \feh .}
\label{fig:ZnFe}
\end{figure}

\begin{figure}
\resizebox{\hsize}{!}{\includegraphics{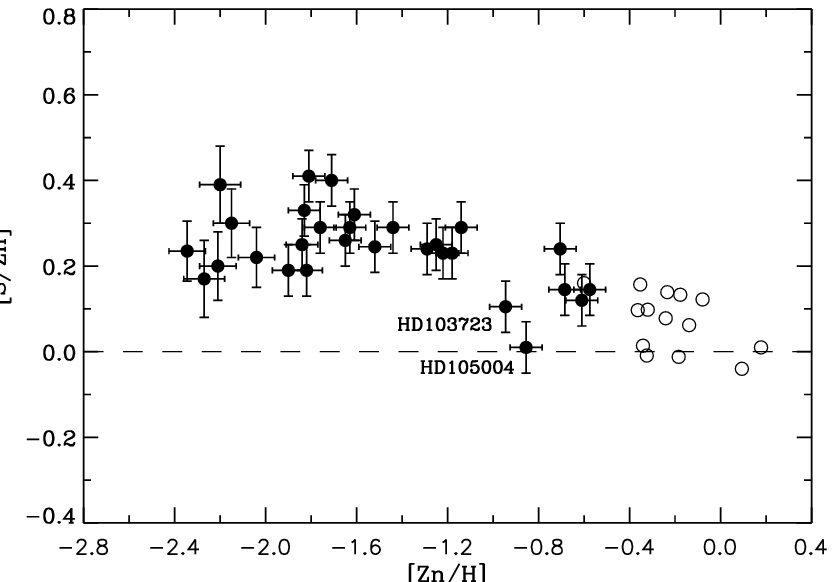}}
\caption{\szn\ vs. \znh . Filled circles: halo stars;
open circles: disk stars from Chen et al. (\cite{chen02}, \cite{chen03}).
The error bars indicate 1-$\sigma$ statistical errors of
\szn\ and \feh .}
\label{fig:SZn}
\end{figure}

Errors in the measured equivalent widths were discussed in
Sect.~\ref{eqw.widths} and found to be about $\pm 0.6$\,m\AA\ for the
\FeII\ and the \ZnI\ lines as well as the
$\lambda \lambda 8694.0, 8694.6$  \SI\ pair,
whereas the error for the 
$\lambda \lambda 9212.9, 9237.5$\,\AA\ \SI\ lines
is $\pm 2.3$m\AA . Adopting these errors, and taking into
account the number of lines observed for a given star
and the contribution from the estimated errors in \teff\
($\pm 70$\,K), \logg\ ($\pm 0.15$\,dex) and 
$\xi_{\rm turb}$ ($\pm 0.2$\kmprs ), we have
estimated individual abundance errors for each star.
These errors are indicated with 1-sigma error bars in
Figs.~\ref{fig:SFe}, \ref{fig:ZnFe}, and \ref{fig:SZn}.
At the lowest metallicities the dominating error contribution
comes from the measurement of the equivalent widths of the
weak absorption lines, whereas errors in the atmospheric parameters
give the largest contribution for the more metal-rich stars.

\begin{table}
\caption[ ]{Changes in derived abundances resulting from the listed
changes in model atmosphere parameters.}
\label{table.change}
\setlength{\tabcolsep}{0.07cm}
\begin{tabular}{lrrrrrr}
\noalign{\smallskip}
\hline
\noalign{\smallskip}
  & $\Delta\ratfeh$ &  $\Delta\ratsh$ & $\Delta\ratznh$ & $\Delta\ratsfe$ & $\Delta\ratznfe$ & $\Delta\ratszn$   \\
\noalign{\smallskip}
\hline
\noalign{\smallskip}
           $\Delta \teff = 100$\,K & 0.01 & $-0.04$ & 0.04 & $-0.05$ &   0.03  & $-0.08$ \\
           $\Delta \logg = 0.2$    & 0.07 &   0.06  & 0.03 & $-0.01$ & $-0.04$ &   0.03  \\
           $\Delta \feh  = 0.2$    & 0.01 &   0.01  & 0.01 &   0.00  &   0.00  &   0.00  \\
$\Delta \xi_{\rm turb} = 0.3$\,km/s$^{\rm a}$ &$-0.06$&$-0.02$&$-0.02$& 0.04 &0.04 & 0.00   \\
\noalign{\smallskip}
\hline
\end{tabular}
$^{\rm a}$ The abundance changes caused by changing $\xi_{\rm turb}$ by  
0.3\,\kmprs\ are maximum values valid for the more metal-rich stars.
\end{table}

\subsection{Systematic abundance errors}
\label{syste.errors}

In this section we discuss systematic abundance errors 
arising from the modelling of the line formation processes using
plane parallel, homogeneous model atmospheres and
the assumption of LTE.

\subsubsection{The influence of stellar granulation}
\label{granulation}
Much attention has recently been paid to the effect on abundance
determinations of three-dimensional (3D)
hydrodynamical modelling of convection and granulation
in stellar atmospheres. As first shown by Asplund et al. (\cite{asplund99}),
there is a very significant surface cooling in metal-poor stars caused by the
near adiabatic cooling of rising and expanding elements coupled with the lack of
radiative heating due to the low line opacity.
The effect on lines formed in the upper atmosphere can be very large.
As an example, Asplund \& Garc\'{\i}a P{\'e}rez (\cite{asplund01})
estimated 3D effects on oxygen abundances derived from OH lines
to be of the order of $-0.6$\,dex
for turnoff stars with $\feh < -2.0$.

\begin{table*}
\begin{center}
\caption[ ]{The effects of 3D hydrodynamical model atmospheres on
the derived sulphur, zinc and iron abundances relative to
those estimated with classical 1D models ($\Delta({\rm log} \epsilon) =
{\rm log} \epsilon(3{\rm D}) - {\rm log} \epsilon(1{\rm D})$).
Both the 1D and 3D abundances have been computed assuming LTE.
It should be noted that the \SI\ 9212.9\,\AA\
and \SI\ 9237.5\,\AA\ lines are very strong ($ W > 100$\,m\AA\ ) at \feh\ $=0.0$
and hence are sensitive to the spectral line broadening. Similarly,
the \SI\ 8694.6\,\AA\ line is very weak ($W < 0.5$\,m\AA\ ) at \feh\ $=-3.0$.
}
\label{table.3D}
\setlength{\tabcolsep}{0.12cm}
\begin{tabular}{ccccccccc}
\noalign{\smallskip}
\hline
\noalign{\smallskip}
 \teff  & \logg & \feh &
\multicolumn{6}{c}{$\Delta({\rm log} \epsilon) =
{\rm log} \epsilon(3{\rm D}) - {\rm log} \epsilon(1{\rm D})$} \\
\cline{4-9}
\noalign{\smallskip}
 $$[K]  &       &      &  \SI\ 8694.6\,\AA\ & \SI\ 9212.9\,\AA\ & \SI\
9237.5\,\AA\ &
                          \ZnI\ 4722.2\,\AA\ & \ZnI\ 4810.5\,\AA\ & \FeII\ \\
\noalign{\smallskip}
\hline
\noalign{\smallskip}
       &      &        &       &      &      &       &      &        \\
  5767 & 4.44 & $-0.0$ & $-0.04$ & $-0.01$ & $-0.01$ & $+0.02$ & $+0.02$ &
$-0.02$ \\
  5822 & 4.44 & $-1.0$ & $+0.02$ & $+0.02$ & $+0.02$ & $+0.01$ & $+0.00$ &
$+0.04$ \\
  5837 & 4.44 & $-2.0$ & $+0.06$ & $+0.13$ & $+0.11$ & $+0.09$ & $+0.09$ &
$+0.10$ \\
  5890 & 4.44 & $-3.0$ & $+0.05$ & $+0.08$ & $+0.08$ & $+0.08$ & $+0.08$ &
$+0.09$ \\
       &      &        &       &      &      &       &      &        \\
  6191 & 4.04 & $-0.0$ & $+0.03$ & $+0.11$ & $+0.09$ & $-0.08$ & $-0.06$ &
$-0.01$ \\
  6180 & 4.04 & $-1.0$ & $+0.04$ & $+0.01$ & $+0.03$ & $-0.02$ & $-0.03$ &
$+0.02$ \\
  6178 & 4.04 & $-2.0$ & $+0.04$ & $+0.09$ & $+0.09$ & $+0.03$ & $+0.03$ &
$+0.06$ \\
  6205 & 4.04 & $-3.0$ & $+0.04$ & $+0.08$ & $+0.07$ & $+0.04$ & $+0.04$ &
$+0.07$ \\
\hline
\end{tabular}
\end{center}
\end{table*}

To investigate the 3D effects on the sulphur and zinc lines used in the present
study a differential study similar to that described by
Asplund \& Garc\'{\i}a P{\'e}rez (\cite{asplund01}) has been carried out. The
results are given in Table~\ref{table.3D} for two sets of model atmospheres with
metallicities of $\feh = 0.0, -1.0, -2.0$ and $-3.0$. The first set has
\teff\ and \logg\ close to the solar values, whereas the second set
represents stars close to the turnoff of the halo population.
For each 3D model atmosphere,
the S and Zn abundances are those reproducing the equivalent widths computed
with 1D MARCS with identical stellar parameters and spectral line data;
for Fe, the results are culled from the identical calculations presented by
Asplund \& Garc\'{\i}a P{\'e}rez (\cite{asplund01}) and correspond to
the average of six \FeII\ lines.
The spectrum syntheses assume \sfe\ $=+0.4$ and \znfe\ $=0.0$ for \feh\ $\le
-1$,
but the results are almost identical for reasonable \sfe\ and \znfe.
It can be seen that the maximum 3D
effect occurs for the most metal-poor stars: the 3D abundance is higher
than the 1D value by 0.07 to 0.13\,dex in the case of the
$\lambda \lambda$9212.9, 9237.5 \SI\ lines. Interestingly, however, there is
a similar 3D effect
on the Fe abundance derived from
\FeII\ lines. Hence, it appears that we are in
the favourable situation that the S/Fe ratio is quite immune to
1D -- 3D effects, with net differences of $\Delta \sfe\ \la 0.05$\,dex
for the stellar parameters typical of our sample.
Similarly, the 3D effects on the Zn/Fe ratio are very small
($\Delta \znfe\ \la 0.05$\,dex).
The small 3D effects and the fact that they in general have the
same sign are a consequence of the similarity in line formation depths
of these \SI , \ZnI\ and \FeII\ lines.

\subsubsection{Non-LTE effects}
\label{non-LTE}

The S and Fe abundances have been derived from lines belonging to
the main ionization stages of the elements, i.e. the neutral stage for
S and the first ionized stage for Fe. Furthermore, the lines correspond
to high excitation levels and are weak in our stars.
Hence, they are formed deep in the stellar atmospheres
where only small departures from LTE are expected. This has been confirmed
in the case of the \FeII\ lines by Th\'evenin \& Idiart (\cite{thevenin99}).
According to Takada-Hidai et al. (\cite{takada02}) non-LTE effects on the
weak, high excitation \SI\ lines at 8694\,\AA\ are estimated to be small ($< 0.05$\,dex).
The good agreement found in Sect~\ref{sulphur.abun} between sulphur abundances
derived from the $\lambda \lambda 8694.0, 8694.6$ pair and the
$\lambda \lambda 9212.9, 9237.5$ lines suggests that  non-LTE effects
on the 9212--9238 \SI\ triplet are also small.

The case of zinc is more difficult. With an ionization potential,
$\chi_{\rm ion} (\ZnI)$ = 9.39\,eV, there are approximately equal
numbers of neutral and ionized zinc atoms at the temperatures and
electron pressures of the line forming regions in the atmospheres  
of our stars. Hence, an over\-ionization of \ZnI\ relative to LTE,
as is the case for \FeI\ (Th\'evenin \& Idiart \cite{thevenin99}),
would lead to an underestimate of the abundance of Zn in the LTE 
analysis of \ZnI\ lines.
However, such an over-ionization effect on \ZnI\  
is likely to be smaller than the equivalent effect 
for \FeI,  since $N_\ZnI \sim N_\ZnII $ 
whereas $N_\FeI << N_\FeII $.
Furthermore, there may be departures from LTE in the excitation
balance of \ZnI , since the lower level from which the lines
originate is the first excited level of \ZnI . 
The discrepancy noted above between the meteoritic Zn abundance 
$\log \epsilon$(Zn) = 4.67 and the photospheric value of 4.57 may 
be due to such non-LTE problems. 
Given the complex structure of the Zn atom,
a thorough study of non-LTE effects on the determination of Zn 
abundances is clearly needed but, to our knowledge, none
has been carried out yet.
Here we simply note that, by analyzing our \ZnI\ lines relative 
to the solar flux spectrum, the problem is somewhat reduced.

Ideally, the non-LTE calculations should
be done in combination with 3D models. This is, however, a very demanding task,
which has only recently become feasible
(e.g. Kiselman \& Nordlund \cite{kiselman95}; 
Kiselman \cite{kiselman97}; Uitenbroek \cite{uitenbroek98};
Asplund et al. \cite{asplund03}).
For complex atoms like Zn, the computationally less demanding 1.5D non-LTE 
problem\footnote{The approximation of treating an ensemble of vertical
atmospheric columns extracted from a 3D model as separate plane-parallel 1D model
atmospheres on which the 1D non-LTE line formation calculations are
performed before averaging, is normally referred to as a 1.5D non-LTE
calculation. Such a procedure ignores all horizontal radiative transfer
in the 3D model atmosphere. While the results are not quantitatively exact, 
they can nevertheless provide a qualitative assessment of the general behaviour of the
full 3D non-LTE problem and yield a useful estimate of the 3D non-LTE effects.} 
would be a good starting point, as recently attempted for \FeI\ 
(Shchukina \& Bueno \cite{shchukina01}).
It is noteworthy that the 3D non-LTE results can be significantly 
different from both the 1D non-LTE and the 3D LTE results.
There are, however, very good reasons
why quite small ($\simlt 0.1$\,dex) 3D non-LTE corrections are expected in
the case of \SI . The \SI\ lines are formed in deep atmospheric layers
where the differences between 1D and 3D model atmospheres are small.
One therefore expects similar non-LTE abundance corrections in 1D
and 3D, as for example appears to be the case for \OI .

\section{Discussion}
\label{discussion}

\subsection{Sulphur as an $\alpha$-capture element}
\label{discussion_sulphur}

As can be seen from Fig.~\ref{fig:SFe}, the halo stars
are distributed around  $\sfe \simeq +0.3$ except for two
deviating stars, \object{HD\,103723} and \object{HD\,105004},
that are known to
have solar-like \alphafe\ ratios as discussed in the next
paragraph. Using
a maximum likelihood program that takes into account individual
errors in both x and y, we obtain a fit for the halo stars
(excluding \object{HD\,103723} and \object{HD\,105004})
\begin{eqnarray}
\sfe = 0.240 \,(\pm 0.024) - 0.042 \,(\pm 0.014) \cdot \feh \nonumber
\end{eqnarray}
with a reduced chi-square $\chi^2_{red}$ = 1.28. If
\object{HD\,103723} and \object{HD\,105004} are included the relation becomes
\begin{eqnarray}
\sfe = 0.153 \,(\pm 0.023) - 0.085 \,(\pm 0.013) \cdot \feh , \nonumber
\end{eqnarray}
but then $\chi^2_{red}$ increases to an unacceptable
high value of 2.3. In any case the slope is much smaller
than the corresponding slope of the fit to the disk star data:
\begin{eqnarray}
\sfe = -0.014 \,(\pm 0.015) - 0.308 \,(\pm 0.040) \cdot \feh . \nonumber
\end{eqnarray}
Hence, there is no tendency in our halo star data for a continuing
strong linear rise of \sfe\ with decreasing \feh , as claimed by
Israelian \& Rebolo (\cite{israelian01}).
At $\feh = -2$ their data suggest [S/Fe] $\simeq$ +0.7, whereas
our relation corresponds to [S/Fe] = +0.32.
We note in this connection that their sample includes
only three stars with $\feh < -1.8$ and none below $\feh = -2.3$,
whereas our sample includes 14 stars in the range $-3.2 < \feh < -1.8$
and seven below $\feh = -2.3$. Furthermore,
the error bars on their sulphur abundance determinations are large
due to the weakness of the \SI\ 8694.6\,\AA\ line; in our
study we circumvent this problem with the 9212.9 and
9237.5\,\AA\ lines which are well suited for
sulphur abundance determinations below $\feh = -2$. Hence, our conclusion
is that there is no need to invoke element production by hypernovae or
very massive supernovae to explain the general behaviour of sulphur.
Our data are in good agreement with traditional
Galactic evolution models with near-instantaneous production
of $\alpha$-elements
by Type II supernovae and delayed production of the iron-peak
elements (e.g. Chiappini et al.  \cite{chiappini99},
Goswami \& Prantzos \cite{goswami00}).

All stars from the
present investigation (plotted with filled circles) have halo kinematics,
including six stars with $-0.9 < \feh < -0.7$
which have Galactic rotational velocities of less than 50\,\kmprs\, i.e.
well below the characteristic rotational velocity of thin disk stars
(225\,\kmprs ) and thick disk stars (175\,\kmprs ). As shown by
Nissen \& Schuster (\cite{nissen97}), 
there is an overlap between halo and thick disk stars in
the metallicity range $-1.0 < \feh < -0.6$. Four of our six stars
in this metallicity range
have enhanced S/Fe ratios like the thick disk stars, but two
(\object{HD\,103723} and \object{HD\,105004}) have a solar S/Fe ratio. 
These two stars also show solar $\alpha$/Fe ratios in 
other $\alpha$-elements, such as O, Mg, Si, Ca, and Ti 
(Nissen \& Schuster \cite{nissen97}).
On the basis of the stars' Galactic orbits, 
Nissen \& Schuster suggested that they may
have been accreted from dwarf galaxies with a chemical evolution
that has proceeded more slowly than in the inner part of our Galaxy,
where the `normal' halo stars formed. 
The fact that S and the 
classical $\alpha$-elements, Mg, Si and Ca, exhibit the same behaviour 
in these `anomalous' stars is a further indication that sulphur belongs 
to this group of elements. 
Thus, there seems to be little ground, on the basis
of our data, for the reservations expressed by
Prochaska et al. (\cite{prochaska00}) concerning the
use of S as an $\alpha$-element in the analysis
of abundance ratios in DLAs.

\subsection{Zinc as a tracer of iron}
\label{discussion_zinc}

Zinc is an interesting element with a number of possible
nucleosynthesis channels: neutron capture ($s$-processing)
in low and intermediate mass stars as well as explosive
burning in Type II and Ia SN (Matteucci et al. \cite{matteucci93}).
Furthermore, zinc is a key element in studies of elemental
abundances of damped Ly$\alpha$ systems, because
it is one of the few elements which in the interstellar medium
are not depleted onto dust and measurements of its interstellar
absorption lines have several other practical advantages 
(Pettini, Boksenberg \& Hunstead \cite{pettini90}).

Interstellar iron, on the other hand, can exhibit large gas depletions,
and it has thus become customary to use Zn as a tracer of Fe
in studies aimed at investigating the chemical
evolution and dust content of galaxies, particularly at high redshifts.
The underlying assumption is that the 
abundances of Zn and Fe vary in lockstep,
as originally found by Sneden, Gratton, \& Crocker (\cite{sneden91}) 
although their data, obtained with 2-3\,m class telescopes, 
exhibited considerable scatter.

Our ignorance of the nucleosynthetic origin of Zn has prompted some
to question the validity of using it as a proxy for Fe. We can
now reassess the validity of this assumption with modern data
such as those presented here. Inspection of Fig.~\ref{fig:ZnFe}
shows that, within the errors, [Zn/Fe] is indeed approximately
solar, at least between [Fe/H]\,=\,0 and $-2$. Over this range the 
mean of the 61 measurements shown in Fig.~\ref{fig:ZnFe}
is [Zn/Fe]\,$ = +0.03 \pm 0.08$~($1 \sigma$).
A similar conclusion was recently reached by 
Mishenina et al. (\cite{mishenina02}) who
published a survey of Zn abundances in 90 disk and halo stars
based on the equivalent widths of the $\lambda \lambda$\,4722.2, 4810.5,
6362.35 \ZnI\ lines in high resolution spectra of dwarf and giant stars
and concluded that their data ``confirm the well-known
fact that the ratio \znfe\ is almost solar at all metallicities''.

\begin{table*}
\begin{center}
\caption[ ]{Interstellar sulphur and zinc measurements in damped Ly$\alpha$ systems.}
\label{dlas.dat}
\setlength{\tabcolsep}{0.175cm}
\begin{tabular}{lrcccccr}
\noalign{\smallskip}
\hline
\noalign{\smallskip}
 QSO & $z_{\rm abs}~~$  & $\log N$\/(\HI) & $\log N$\/(\SII)& $\log N$\/(\ZnII) & [Zn/H]$^{a}$ & [S/Zn]$^{b}$  & Ref.$^{c}$ \\
     &                  &  (cm$^{-2}$)    &  (cm$^{-2}$)    & (cm$^{-2}$)       &              &               & \\
\hline
\noalign{\smallskip}
\object{Q0000$-$2620}    & 3.3901   & $21.41 \pm 0.08$ & $14.70 \pm 0.03$  & $12.01 \pm 0.05$ & $-2.07$ &   $+0.16$ &   1,\,2 \\
\object{Q0013$-$004}     & 1.97296  & $20.83 \pm 0.05$ & $15.28 \pm 0.03$  & $12.74 \pm 0.04$ & $-0.76$ &   $+0.01$ &   3 \\
\object{Q0100$+$130}     & 2.3090   & $21.40 \pm 0.05$ & $15.13 \pm 0.05$  & $12.45 \pm 0.10$ & $-1.62$ &   $+0.15$ &   2,\,4 \\
\object{Q0407$-$4410}    & 2.5505   & $21.13 \pm 0.10$ & $14.82 \pm 0.06$  & $12.44 \pm 0.05$ & $-1.36$ &   $-0.15$ &   5 \\
\object{Q0407$-$4410}    & 2.5950   & $21.09 \pm 0.10$ & $15.19 \pm 0.05$  & $12.68 \pm 0.02$ & $-1.08$ &   $-0.02$ &   5 \\
\object{Q0551$-$366}     & 1.96221  & $20.50 \pm 0.08$ & $15.38 \pm 0.11$  & $13.02 \pm 0.05$ & $-0.15$ &   $-0.17$ &   6 \\
\object{0841+1256}       & 2.3745   & $21.00 \pm 0.10$ & $14.77 \pm 0.03$  & $12.20 \pm 0.05$ & $-1.47$ &   $+0.04$ &   7 \\
\object{HE2243$-$6031}   & 2.33000  & $20.67 \pm 0.02$ & $14.88 \pm 0.01$  & $12.22 \pm 0.03$ & $-1.12$ &   $+0.13$ &   8 \\
\object{B2314$-$409}     & 1.8573   & $20.90 \pm 0.10$ & $15.10 \pm 0.05$  & $12.52 \pm 0.03$ & $-1.05$ &   $+0.05$ &   9 \\
\object{Q2343$+$125}     & 2.4313   & $20.35 \pm 0.05$ & $14.71 \pm 0.08$  & $12.45 \pm 0.06$ & $-0.57$ &   $-0.27$ &   3,\,10 \\
\noalign{\smallskip}
\hline
\end{tabular}
\smallskip

\begin{minipage}{160mm}
\smallskip
\item[$^{\rm a}$] [$\log$ (Zn/H)$_{\rm DLA} - \log$ (Zn/H)$_{\odot}$],
where $\log$ (Zn/H)$_{\odot} = -7.33$ (Grevesse \& Sauval \cite{grevesse98}).

\item[$^{\rm b}$] [$\log$ (S/Zn)$_{\rm DLA} - \log$ (S/Zn)$_{\odot}$],
where $\log$ (S/Zn)$_{\odot} = 2.53$ (Grevesse \& Sauval \cite{grevesse98}).

\item[$^{\rm c}$] References---1: Molaro et al. (\cite{molaro00}); 
2: Lu, Sargent, \& Barlow (\cite{lu98});
3: Petitjean, Srianand \& Ledoux (\cite{petitjean02})
4: Prochaska et al. (\cite{prochaska01});
5: Lopez \& Ellison (\cite{lopez03});
6: Ledoux, Srianand, \& Petitjean (\cite{ledoux02});
7: Cent\'{u}rion et al. (\cite{centurion03});
8: Lopez et al. (\cite{lopez02});
9: Ellison \& Lopez (\cite{ellison01});
10: Dessauges-Zavadsky, Prochaska, \& D'Odorico (\cite{dessauges02}).

\end{minipage}
\end{center}
\end{table*}

Looking more closely at the data in Fig.~\ref{fig:ZnFe},
there may be hints of subtle
trends which, if confirmed by larger samples, could provide
clues to the nucleosynthetic origin of Zn.
For example, Prochaska et al. (\cite{prochaska00})
claimed that in thick disk stars $\znfe \simeq +0.1$.
This mild overabundance of Zn relative to Fe in thick disk stars
is also present in the Mishenina et al. (\cite{mishenina02}) sample, once the
stars are separated on the basis of their kinematics into 
halo, thick and thin disk populations, as recently pointed out by
Nissen (\cite{nissen03}).
Furthermore, there seems to be a gradient in \znfe\ as a function of
\feh\ for the halo stars in the  Mishenina et al. sample
with the highest values of \znfe\ being measured in the most metal-poor stars. 
The few stars in the present study with $\feh < -2.0$ show a similar
effect, with a mean $\znfe \simeq +0.1$ (see Fig.~\ref{fig:ZnFe}).
On the other hand, before making too much of these trends, it is
important to bear in mind that there may well be 
systematic errors at the $\pm 0.10$\,dex level in the 
determinations of \znfe,  as discussed in Sect.~\ref{zinc.abun}.
Clearly, \znfe\ in halo and disk stars should be studied further, 
both observationally and in terms of Galactic chemical evolution models.

Notwithstanding these uncertainties, it can be seen from 
Fig.~\ref{fig:SZn}
that the \szn\ ratio shows the same trend vs. metallicity as the
classical [$\alpha$/Fe] trend, i.e. a plateau at
$\szn \simeq +0.25$ until $\znh \simeq -1.0$ and then
a decline of \szn\ to the solar ratio at $\znh \simeq 0$. 
Complications may arise, however, in the transition region between 
the halo and the disk where the `low'-$\alpha$ stars identified to date
occur. More measurements in this metallicity regime would be highly
desirable.

\subsection{Sulphur and zinc in damped Ly$\alpha$ systems}
\label{dlas}

Armed with the results of Fig.~\ref{fig:SZn},
we are now in a position to examine the \szn\
ratio in DLAs and compare it to the values measured
in Galactic stars. Despite the large database of
abundance measurements in DLAs accumulated over the last
few years, there are still relatively few 
determinations of \szn\ in DLAs. The reasons are two-fold.
First, the rest wavelengths of the \SII\ triplet 
$\lambda\lambda 1250, 1253, 1259$
are so close to the wavelength of Ly$\alpha$ (1215.67\,\AA)
that the \SII\ lines often fall within the 
Ly$\alpha$ forest, where blending can be a problem.
Second, with rest wavelengths $\lambda \lambda 2026, 2062$,
the \ZnII\ doublet lines are separated by
$ \sim (1+z_{\rm abs}) \times 800$\,\AA\
from the \SII\ triplet; thus, 
at the redshifts $z_{\rm abs} = 2-3$
of most DLAs, the two sets of spectral features fall in widely
separated regions of the optical spectrum which may not
be covered simultaneously by some instruments.
For this reason, most measurements of
the \szn\ ratio in DLAs have become available
only recently thanks to the wide spectral coverage and high
efficiency at blue and red wavelengths of the
VLT/UVES combination.

\begin{figure}
\vspace{-3.75cm}
\hspace{-1.0cm}
\resizebox{10.3cm}{!}{\includegraphics{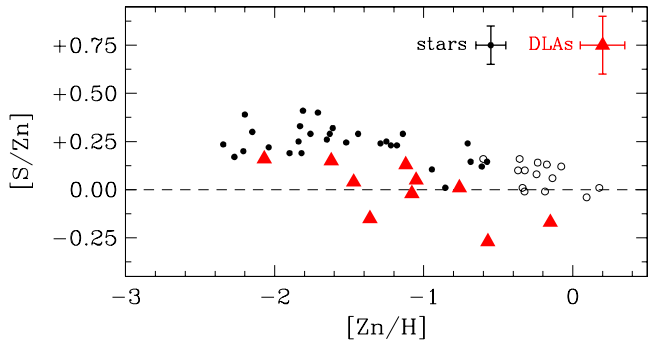}}
\vspace{-5cm}
\caption{\szn\ vs. \znh\ in Milky Way stars and 
damped Ly$\alpha$ systems. Typical errors are shown.}
\label{fig:dlas}
\end{figure}

In Table~\ref{dlas.dat} we have collated 
from the literature
all measurements of \szn\ in DLAs from spectra
obtained with 8-10\,m class telescopes; 
references to the original works are
given in the last column of the Table.
The total sample consists of ten DLAs at redshifts
$z_{\rm abs} = 1.86 - 3.39$; their values of 
\szn\ vs. \znh\ are compared with the stellar
measurements in Fig.~\ref{fig:dlas}.
The behaviour of the \szn\ ratio in DLAs is evidently
not the same as that seen in Galactic stars.
Taken together, the ten DLAs considered here do not
show evidence for an $\alpha$-element enhancement;
the mean and standard deviation for the sample
are \szn\,$=-0.01 \pm 0.15$. Considering only
the seven DLAs with \znh\,$< -1$, we find 
\szn\,$=+0.05 \pm 0.11$.

It is hard to identify systematic effects
which may be the cause of this apparent offset between DLAs
and Galactic halo stars. 
The column density errors listed in Table~\ref{dlas.dat},
which in turn translate to the typical abundance 
errors illustrated in Fig.~\ref{fig:dlas}, are those
quoted by the authors of the original papers, as
referenced in the Table. These errors are
the $1 \sigma$ uncertainties returned by the
absorption line fitting computer codes and 
generally reflect the random errors in the line 
equivalent widths. They are probably
underestimates of the true uncertainties (see, for example, the 
discussion of this point by Kirkman et al. \cite{kirkman03}),
but this should result in an
increased scatter of the data points in Fig.~\ref{fig:dlas},
rather than a systematic offset.
Generally, the \SII\ triplet lines are
stronger than the \ZnII\ doublet, so that saturation may be
an issue. However, with three absorption lines available,
it is normally possible to assess the degree of saturation
reliably, unless the distribution of absorber properties is
markedly irregular (Jenkins \cite{jenkins86}); the line fitting 
programs used in the original analyses of the data in 
Table~\ref{dlas.dat} are well up to this task.
Similar considerations apply to resolving the 
\ZnII\ $\lambda\lambda 2026.1, 2062.7$
doublet lines from the nearby \MgI\ $\lambda 2026.5$ and 
\CrII\ $\lambda 2062.2$ lines.

Turning from ion column densities to element abundances,
it is also difficult to find reasons why the
abundance of S should have been systematically underestimated,
or that of Zn overestimated.
Both \SII\ and \ZnII\ are the
major ionization stages of their respective elements
in \HI\ regions, and corrections for unobserved ion stages
are expected to be unimportant (e.g. Vladilo et al. \cite{vladilo01}).
Neither S nor Zn show much affinity for dust
and the problem is further lessened in DLAs which generally
show only mild depletions of even the refractory elements
(Pettini et al. \cite{pettini97}). 
Errors in the solar abundance scale do not seem a plausible explanation
for the difference either. Recall that while the stellar
abundances are derived differentially relative to the Sun,
those in DLAs have to be referred to a solar scale;
here we have adopted the meteoritic abundances of S and Zn from the
compilation by Grevesse \& Sauval (\cite{grevesse98}).
However, as discussed earlier (Sect. 4.2 and 4.3), 
the \SI\ lines used in the present study
give a solar photospheric abundance of S which is
the same as the meteoritic one, and while for Zn
there is a 0.10\,dex offset, this offset is in the 
wrong direction for reconciling stellar and DLA data.

In summary, on the basis of our current knowledge of S and Zn
in the local interstellar medium, 
the values given in Table~\ref{dlas.dat} 
should reflect the true interstellar abundances of
these two elements in the high redshift galaxies giving rise to 
the damped Ly$\alpha$ systems. 
Thus, we are led to conclude that the difference 
between Milky Way stars and DLAs is probably real,
confirming the earlier analysis by
Centuri\'{o}n et al. (\cite{centurion00})
which was based on noisier 4-m data.
A similar conclusion has also been reached by several
studies which targeted depleted elements
(chiefly Si and Fe) and then attempted to correct
for the fractions in dust (e.g. Vladilo \cite{vladilo02} 
and references therein), although the method
used here is clearly more direct, as it does not rely on the
accuracy of the dust corrections.

Interpreted within standard chemical evolution models
(e.g. Calura, Matteucci, \& Vladilo \cite{calura03}),
the lack of $\alpha$-element enhancement in DLAs
may be taken as evidence for low and 
intermittent rates of star formation over their past
history (relative to the time when we observe them).
Put simply, in this scenario the overall metallicity grows 
only slowly with time and the
iron-peak elements released by Type Ia supernovae
can `catch-up' with the overall chemical enrichment
during quiescent periods between isolated bursts of
star formation.
Such a mode of star formation is seen locally in
dwarf irregular and dwarf spheroidal galaxies whose stars also do not show
an $\alpha$-element enhancement at low
metallicities (Carigi, Hernandez, \& Gilmore \cite{carigi02}; 
Venn et al. \cite{venn03}; Shetrone et al. \cite{shetrone03};
Tolstoy et al. \cite{tolstoy03}), and is indeed envisaged
in theoretical models of DLA galaxies 
(e.g. Mo, Mao, \& White \cite{mo98}).

Nevertheless, it is somewhat remarkable that 
we have not found even a single DLA 
with a high \szn\ ratio in our (albeit small) sample.
At lower redshifts ($z < 1$), DLA galaxies are a heterogeneous
group, exhibiting a variety of morphologies and surface brightnesses
(Boissier, P{\' e}roux, \& Pettini \cite{boissier03}); if this is also
the case at high $z$, we may have expected a wider range of
\szn\ values. Possibly the issue is still 
clouded by small number statistics, or there may be more
fundamental differences between high- and low-redshift DLAs.
The findings by Kanekar \& Chengalur (\cite{kanekar03}) may be 
relevant here. These authors derived estimates
of the spin temperature $T_{\rm s}$ in 24 DLAs, 11 of which
(all at $z_{\rm abs} < 1$) have optical identifications.
Kanekar \& Chengalur (\cite{kanekar03}) find that all DLAs with 
high values of spin temperature ($T_{\rm s} \simgt 1000$\,K) 
are identified with dwarf or low surface brightness galaxies,
while DLAs with low values of  $T_{\rm s}$ are invariably
associated with large, luminous galaxies.
Furthermore, low $z$ DLAs exhibit both high and low values
of $T_{\rm s}$, while high redshift ($z_{\rm abs} \simgt 2$) 
DLAs have preferentially high spin temperatures.
The lack of a significant $\alpha$-element
enhancement in the DLAs considered here 
can be understood within the
picture put forward by Kanekar \& Chengalur (\cite{kanekar03}), 
if dwarf and low surface brightness galaxies
dominate the cross-section for DLA absorption at high
redshift, and if these types of galaxies 
generally tend to have low and
intermittent rates of star formation.

Centuri\'{o}n et al. (\cite{centurion00}) and Vladilo (\cite{vladilo02})
suggested that there may be a mild trend of decreasing 
\szn\ (and more generally [$\alpha$/Fe])
with increasing metallicity in DLAs. While the DLA
data shown in Fig.~\ref{fig:dlas} are not inconsistent
with such a possibility, 
the number of reliable S and Zn abundance measurements needs
to be considerably larger than the current sample before
the reality of such a trend can be assessed statistically.

\section{Summary and conclusions}
\label{summary}
One of the motivations for carrying out the present survey of 
S and Zn abundances in Galactic stars is the important role
played by these two elements in deciphering the chemical
enrichment and star formation histories of damped Ly$\alpha$
systems. Up to now measurements of the sulphur abundance in Galactic 
stars have been sparse and doubts have been raised as to whether
S is a typical $\alpha$-capture element formed in Type II supernovae 
(Israelian \& Rebolo \cite{israelian01};  Takada-Hidai et al.
\cite{takada02}). The lack of data for sulphur, coupled with 
insufficient knowledge of the nucleosynthetic channels for the 
production of Zn, has led some (e.g. Prochaska et al. \cite{prochaska00})
to question the validity of using these elements, and in particular
their ratio, in the interpretation of the abundance patterns seen
in DLAs.

The new data presented here have clarified the situation as regards
Milky Way stars. 
By targeting the \SI\ $\lambda\lambda 9212.9, 9228.1, 9237.5$
triplet we have overcome the limitations of most earlier studies 
and probed the abundance of sulphur with higher precision
and to lower metallicities than had been possible previously.
We find that the trend of [S/Fe] as a function of \feh\
is very similar to those of other typical $\alpha$-capture
elements, Mg, Si and Ca. [S/Fe] is nearly constant at a level of
$\sfe \simeq +0.3$\,dex
in the metallicity range $-3.2 < \feh < -1.2$, starts to decrease 
at $\feh \simeq -0.7$, and reaches a solar ratio at $\feh \sim 0$.
In the halo-disk transition region, at metallicities $-1.2 < \feh < -0.7$,
there is significant scatter in the values of \sfe. 
More stars in this interval 
should be studied to look for possible correlations between stellar
kinematics and \sfe . Precise abundance analyses by Fuhrmann
(\cite{fuhrmann98}); Gratton et al. (\cite{gratton00}); and
Feltzing et al. (\cite{feltzing03}) have shown a clear separation
in \mgfe\ between thin and thick disk stars and it will be important
to establish if this is also the case for \sfe.

As far as zinc is concerned, we confirm the results of
several earlier surveys which have shown that, to a first
approximation, Zn tracks Fe over three orders of magnitude
in \feh. There is some evidence for a small 
overabundance of Zn ($\znfe \simeq +0.10$) for metal-poor disk stars
and halo stars with $\feh < -2$. However, there may be systematic
errors in \znfe\ at a level of $\pm 0.1$\,dex due to the difficulty
in analyzing the $\lambda \lambda 4722.2, 4810.5$ \ZnI\
lines in the solar spectrum.
Non-LTE effects on the derived zinc
abundances are also a potential problem that should be addressed.
New studies of the nucleosynthesis of zinc
in supernovae and by the $s$-process in AGB stars would
help in understanding the reasons why Zn behaves like an Fe-peak
element and the origin of the 0.1\,dex offset at low metallicities.

When we turn to damped Ly$\alpha$ systems, however, the Galactic
pattern of S and Zn abundances does not seem to apply. The sample
is still small (accurate measurements of [S/Zn] are available 
for only ten DLAs) and there is scatter in the data but, 
taken at face value, there is no obvious $\alpha$-element
enhancement in DLAs. We can not identify any systematic effect
in the data nor in their analysis which would mask such an 
enhancement, if it were there, and conclude that its absence 
is probably real. Presumably, it is indicative of a bursting
history of star formation 
in the galaxies giving rise to damped
systems. 
However, it is important to realise that 
a variety of recent observations suggest
that the $\alpha$-enhancement
exhibited by metal-poor stars of the Milky Way may
in fact be the exception rather than the rule---it is generally not seen
in dwarf spheroidal and dwarf irregular galaxies (Venn et al. \cite{venn03};
Shetrone et al. \cite{shetrone03}; Tolstoy et al. \cite{tolstoy03};
Aloisi et al. \cite{aloisi03}), in 
old stars of the Large Magellanic Cloud (Hill et al. \cite{hill00}),
in DLAs (this paper), in some Galactic halo stars with 
large orbits (Nissen \& Schuster \cite{nissen97}),
and in the globular cluster \object{Pal\,12} which may originally
have been part of the \object{Sgr\,dSph} galaxy (Cohen \cite{cohen03}).
The challenge now is to incorporate
this rapidly growing set of abundance measurements into a
comprehensive picture of the chemical evolution of galaxies.

\begin{acknowledgements}
The ESO staff at Paranal is thanked for carrying out the VLT/UVES
service observations in a very competent way. In particular we
acknowledge important advice on the observing procedure from Vanessa Hill
and Francesca Primas.
PEN acknowledges support from the Danish Natural Science Research Council (grant 
21-01-0523). MA has been supported by grants from the Swedish Natural Science
Research Council (grants F990/1999 and R521-880/2000), the Swedish 
Royal Academy of Sciences, the G\"oran Gustafsson Foundation and the
Australian Research Council (grant DP0342613). This research has made
use of the SIMBAD database, operated at CDS, Strasbourg, France.
\end{acknowledgements}

{}

\end{document}